\pdfoutput=1
\documentclass[aps,prl,10pt,twocolumn,superscriptaddress,nobibnotes,longbibliography]{revtex4-2}
\usepackage{amsmath,amssymb}
\usepackage{dsfont}
\usepackage{graphicx}
\graphicspath{{./}{Science_submission/}}
\usepackage[hidelinks,hypertexnames=false]{hyperref}

\makeatletter
\newcommand{\SItocHide}{%
  \let\SItoc@section\l@section
  \let\SItoc@subsection\l@subsection
  \let\l@section\@gobbletwo
  \let\l@subsection\@gobbletwo}
\newcommand{\SItocShow}{%
  \let\l@section\SItoc@section
  \let\l@subsection\SItoc@subsection}

\newenvironment{supplementaryfigure}{%
  \par\addvspace{\intextsep}\noindent
  \begin{minipage}{\linewidth}
    \centering
    \def\@captype{figure}%
    \setkeys{Gin}{width=\linewidth,height=.78\textheight,keepaspectratio}%
}{%
  \end{minipage}\par\addvspace{\intextsep}%
}
\makeatother

\newcommand{\manuscripttitle}{Non-integer high harmonic generation from many-body nonadiabatic gating}

\begin{document}

\title{\manuscripttitle}

\author{Mingyao~Guo}
\email[Corresponding author. Email: ]{mguo@caltech.edu}
\affiliation{Department of Physics, California Institute of Technology, Pasadena, CA, USA.}
\affiliation{Institute for Quantum Information and Matter, California Institute of Technology, Pasadena, CA, USA.}

\author{Luting~Wang}
\affiliation{Department of Chemistry, Emory University, Atlanta, GA, USA.}

\author{Murali~Suganya}
\affiliation{Institute of Physics, Academia Sinica, Taipei, Taiwan.}

\author{Biqiong~Yu}
\affiliation{School of Physics and Astronomy, University of Minnesota, Minneapolis, MN, USA.}

\author{Martin~Greven}
\affiliation{School of Physics and Astronomy, University of Minnesota, Minneapolis, MN, USA.}

\author{Sritharan~Krishnamoorthi}
\affiliation{Institute of Physics, Academia Sinica, Taipei, Taiwan.}

\author{Raman~Sankar}
\affiliation{Institute of Physics, Academia Sinica, Taipei, Taiwan.}

\author{Yao~Wang}
\affiliation{Department of Chemistry, Emory University, Atlanta, GA, USA.}

\author{David~Hsieh}
\email[Corresponding author. Email: ]{dhsieh@caltech.edu}
\affiliation{Department of Physics, California Institute of Technology, Pasadena, CA, USA.}
\affiliation{Institute for Quantum Information and Matter, California Institute of Technology, Pasadena, CA, USA.}

\begin{abstract}
High harmonic generation (HHG) in solids typically produces a rigid frequency comb locked to exact integer multiples of the drive photon energy. In strongly correlated materials, however, the light field can dynamically transform the electronic state on the timescale of harmonic generation itself, potentially violating this rule. Here, we report giant blueshifted, non-integer HHG in a cuprate Mott insulator Sr$_2$CuO$_3$ produced by mid-infrared excitations. Concurrent transient reflectivity measurements and time-dependent exact-diagonalization simulations of a driven Hubbard model support a many-body temporal gating mechanism that suppresses light emission on the trailing edge of the pulse, thereby blueshifting radiated harmonics in the frequency domain. Our findings establish HHG emission energy as a sensitive probe of quantum many-body dynamics and pave the way for exploration and manipulation of correlated electronic phases of matter with strong-field spectroscopy.
\end{abstract}
\maketitle
\addtocontents{toc}{\protect\SItocHide}

In condensed matter, strong correlations can overturn rules that hold for independent electrons --- for instance, converting the integer quantum Hall plateaus \cite{klitzing1980new} into a fractional hierarchy \cite{tsui1982two}. Light-matter interaction obeys symmetry-based selection rules, governing processes such as dipole and multipole transitions, inelastic scattering \cite{devereaux2007inelastic,ament2011resonant}, and nonlinear optics \cite{orenstein2021topology,neufeld2026light}, yet how many-body interactions rewrite them, and the information encoded in such deviations, remain a topic of ongoing investigation. High harmonic generation (HHG) represents a canonical example of an extremely nonlinear optical phenomenon in which an intense, low-frequency electromagnetic wave is upconverted into a broad spectrum of high-frequency radiation \cite{ferray1988multiple,ghimire2011observation}. In rigid-band insulators, this process originates from sub-cycle dynamics of charge carriers, commonly understood through a single-particle framework involving intraband acceleration along Bloch bands and interband polarization between electrons and holes \cite{ghimire2019high,goulielmakis2022high}. Typically, the valence band remains far from depleted, and the coherent nonlinear current can be regenerated on successive optical cycles and sustained throughout the pulse envelope as charge excitations propagate independently \cite{sommer2016attosecond,hohenleutner2015real,uchida2024high}. The emission thus repeats with the periodicity of the laser field, producing a rigid comb of sharp peaks at integer multiples of the fundamental carrier frequency.

Mott insulators, in which electronic correlations bind the constituent electrons into a fragile, highly entangled many-body state, represent a qualitative departure from this picture. Numerical simulations show that HHG in these systems relies crucially on the quantum coherence between the excited-state components of the evolving wavefunction and the ground state \cite{silva2018high,murakami2018high}. Under strong-field excitation, the electronic state can dynamically reconstruct far out of equilibrium on sub-cycle timescales \cite{wang2017producing,murakami2025photoinduced}, modifying both the phase and temporal profile of harmonic emission. The resulting interplay between evolving radiation phases and temporal modulation of emission can potentially generate non-integer harmonic spectra that carry signatures of ultrafast quantum many-body phenomena.

The cuprate Mott insulator Sr$_2$CuO$_3$ serves as an ideal platform to explore these strongly driven dynamics. Within its Cu-O chains, Coulomb repulsion splits the Cu $d$-manifold into lower and upper Hubbard bands, opening a robust Mott gap that suppresses trivial laser heating effects \cite{kim2008bound}. Unlike in conventional band insulators, the driving field simultaneously generates charge excitations responsible for HHG and progressively destabilizes the correlated state that supports their coherent emission. In principle, at sufficiently high doublon-holon density, the resulting light-induced electronic state reconstruction can suppress sustained harmonic generation, truncating the emission window and detuning the harmonics from their nominal integer orders. Recent work on Sr$_2$CuO$_3$ revealed modest harmonic redshifts associated with melting of the Mott state, including the effect of gap narrowing \cite{ikeda2025high}. However, whether the theoretically predicted temporal gate \cite{silva2018high,orthodoxou2021high,ikeda2025high} can produce a dominant nonadiabatic spectral shift remains experimentally unresolved  (see discussions in Supplementary Information section~\ref{sec:sbe gap reduction}). Here, we unveil this phenomenon in Sr$_2$CuO$_3$ using longer-wavelength mid-infrared (MIR) driving at substantially higher excitation strength. In this regime, field-driven reorganization of the correlated electronic state suppresses trailing-edge radiation and produces giant, order-dependent blueshifts of the emission energy, rendering the harmonics non-integer. This advances HHG from reflecting transient electronic structure to time-resolving the sub-cycle evolution of the many-body emitter itself.

\section*{Non-integer HHG from a Mott insulator}

\begin{figure}[!htbp]
\centering
\includegraphics[width=\columnwidth]{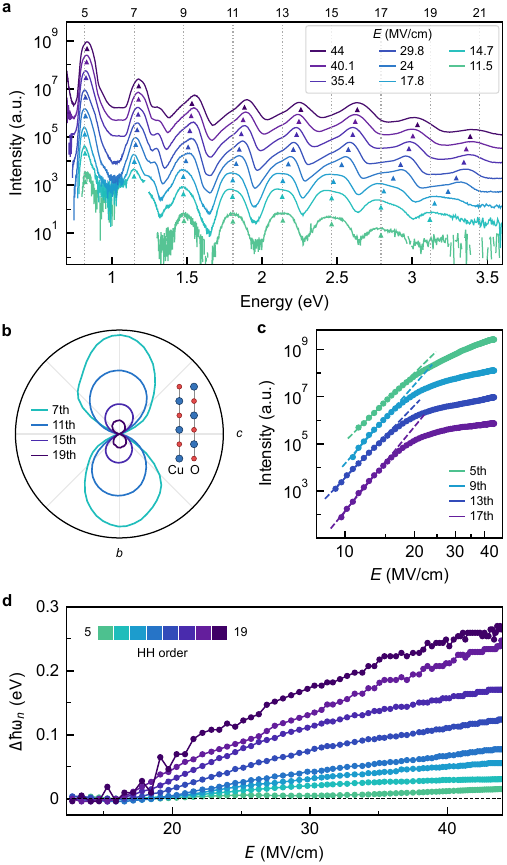}
\caption{\label{fig:m1} Observation of blueshifted non-integer HHG in Sr$_2$CuO$_3$. (a) HH emission spectra at various field strengths driven by 7550~nm MIR pulses along the Cu-O chain direction, vertically offset for clarity. The grey dotted lines denote the expected emission energies of odd-order HHs, and the triangular markers denote the measured harmonic central energies (Supplementary Information section~\ref{sec:HHG profile}). (b) Incident MIR polarization dependence of the HH emission. The inset shows the schematic of CuO chains along the $b$ axis. (c) Integrated HH intensity as a function of peak electric field. The dashed lines are power-law fits to signals below 15~MV/cm, giving exponents $\alpha_5=4.8$, $\alpha_9=5.6$, $\alpha_{13}=5.1$, and $\alpha_{17}=6.0$ in $I_n\propto E^{2\alpha_n}$. (d) HH central emission energies relative to odd multiples of the drive frequency as a function of field strength.}
\end{figure}

We irradiated single crystals of Sr$_2$CuO$_3$ with intense MIR pulses at a central wavelength of 7550~nm (0.164~eV). This low excitation energy is nearly an order of magnitude smaller than the $\sim$1.5~eV Mott-gap \cite{kim2008bound,ikeda2025high,baykusheva2026quantum}, heavily suppressing multi-photon absorption and enabling application of atomically strong electric fields without dielectric breakdown. The HHG spectra acquired in reflection geometry are shown in Fig.~\ref{fig:m1}a, revealing clear odd-order harmonics spanning the infrared, visible, and near-ultraviolet spectral ranges, with harmonic orders up to $n=19$ (even orders are forbidden due to crystalline inversion symmetry \cite{ami1995magnetic}), limited by the efficiency of the detector and reflection optics (\hyperref[sec:experimental setup]{Methods}). Strikingly, as the peak electric field increases, the central frequencies of high harmonics (HHs) do not remain anchored to odd integer multiples of the MIR photon energy, indicated by the vertical dotted lines in Fig.~\ref{fig:m1}a. Instead, they exhibit a pronounced shift that systematically increases with field strength and harmonic order. In addition, the non-integer HHs become increasingly asymmetric relative to the peak, with a broader tail on their low-energy side (Supplementary Information section~\ref{sec:HHG profile}).

Figures~\ref{fig:m1}b and~\ref{sfig:polarization dependence} detail the polarization dependence of the harmonic yield, which displays a clear two-lobe pattern that is maximized when the drive field is parallel to the Cu-O chains, consistent with nonlinear current along the chain direction. Furthermore, the field dependence of the integrated harmonic intensities exhibits a clear deviation from perturbative $I_n\propto E^{2n}$ scaling (Figs.~\ref{fig:m1}c and~\ref{sfig:field dependence}), confirming the non-perturbative nature of the emission.

\begin{figure*}[!htbp]
\centering
\includegraphics[width=\textwidth]{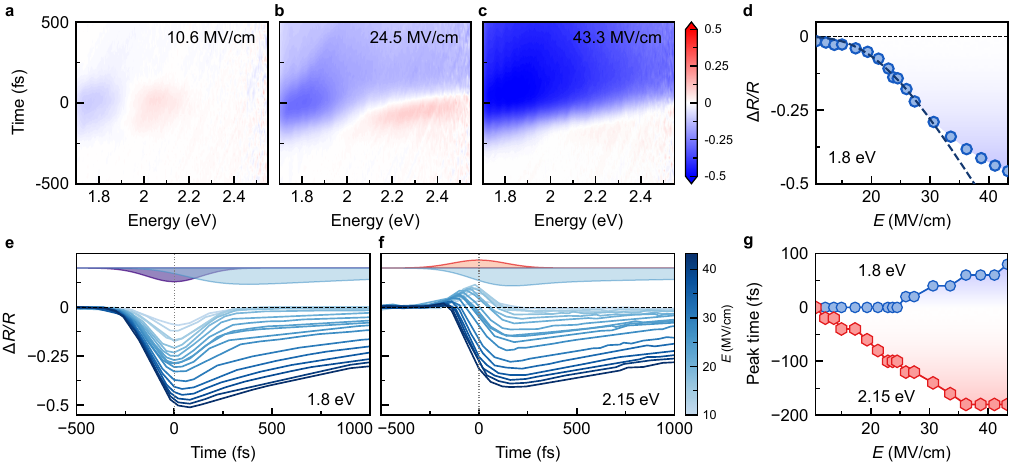}
\caption{\label{fig:m2} Ultrafast electronic dynamics and tunneling-driven electronic state reconstruction. (a)–(c) Maps of transient differential reflectivity spectra probed by a broadband supercontinuum at low (a), intermediate (b), and high (c) 7550~nm driving field strengths. (d) Reflectivity transients extracted at 1.8~eV and 250~fs time delay. The dashed line is a fit to the tunneling formula at low fields. (e), (f) Field-dependent temporal cuts of $\Delta R/R$ at 1.8~eV (e) and 2.15~eV (f). The purple/red and blue shades at the top of panel (e)/(f) schematically show the decomposition of the signal into coherent and incoherent responses, respectively. (g) Time delay at which the absolute differential reflectivity transients peak at 1.8~eV (blue circles) and 2.15~eV (red hexagons) as a function of field strength. The full dataset of $\Delta R/R$ spectra is shown in Supplementary Information section~\ref{sec:trR}.}
\end{figure*}

Quantifying the field dependence of the observed giant blueshift, we find that it sets in above $\sim$16~MV/cm and grows with harmonic order, reaching 0.27~eV at the 19th harmonic without clear saturation (Fig.~\ref{fig:m1}d). The blueshift in Sr$_2$CuO$_3$ is inconsistent with the documented extrinsic origins of frequency shifts in solid-state HHG. First, cross-phase modulation can modify the emitted HHs during propagation \cite{xia2018nonlinear,floss2018ab,yamada2023propagation,koll2025extreme}; however, our reflection geometry restricts the interaction volume to the nanometer-scale skin depth and coherence length, effectively preventing the accumulation of such phase shifts (Supplementary Information section~\ref{sec:propagation}). Second, spectral shifts can arise from carrier-envelope phase effects \cite{schmid2021tunable,graml2023influence}, which are excluded here by our non-phase-stable driving pulses (\hyperref[sec:experimental setup]{Methods}). Finally, plasma and Kerr self-phase modulation effects could reshape the penetrated field \cite{van2023transient,jessen2025interplay}. However, the reflected MIR spectra show no appreciable shift over the deployed excitation condition (Fig.~\ref{sfig:mir reflectivity}), thereby precluding this mechanism. Control measurements on band semiconductors including GaAs, InP, and Ge using the same optical setup yielded integer harmonics despite carrier generation, in accord with previous reports \cite{xia2021high,sekiguchi2022polarization,sekiguchi2023enhancing} (Supplementary Information section~\ref{sec:benchmark}). Notably, we also observed similar blueshifted non-integer HHG under the same experimental conditions in a quasi-two-dimensional Mott-insulating cuprate Sr$_2$CuO$_2$Cl$_2$ (Fig.~\ref{sfig:Sr2CuO2Cl2_HHG}), implying that the underlying mechanism may be generic to Mott insulators. Remarkably, the emission energies do not change with temperature (Supplementary Information section \ref{sec:HHG T dep}), ruling out a thermal origin and establishing the blueshift as an intrinsic property of the strongly driven correlated electronic state.

\section*{Tunneling-driven electronic state reconstruction}

The role of many-body correlations in non-integer HHG can be further elucidated in the time domain. To this end, we employ transient reflectivity measurements to investigate the charge dynamics during this strong-field process. A broadband visible probe pulse is obtained through supercontinuum generation in a sapphire crystal (\hyperref[sec:experimental setup]{Methods}), whose energy spans $1.7\sim2.55$~eV, allowing us to track the evolution of charge excitations under the same MIR drive. Figure~\ref{fig:m2}a displays the transient differential reflectivity map ($\Delta R/R$) at a low field strength (10.6~MV/cm). The response is predominantly coherent, characterized by a signal that almost strictly follows the pump-probe overlap without leaving a residual change after the pump pulse is gone. This indicates that the optical response arises from an instantaneous and non-dissipative distortion of the Mott state. Concomitantly, HHs do not exhibit any shift from their expected integer emission energies (Fig.~\ref{fig:m1}d).

\begin{figure*}[!htbp]
\centering
\includegraphics[width=\textwidth]{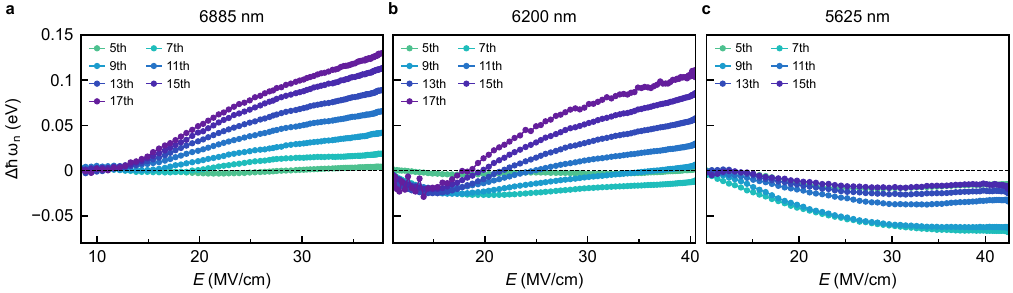}
\caption{\label{fig:m3} Drive-wavelength dependence and crossover of harmonic energy shifts. (a)–(c) Harmonic central emission energy shifts extracted as a function of peak field strength for MIR wavelengths of 6885~nm (a), 6200~nm (b), and 5625~nm (c).}
\end{figure*}

As the field strength enters the regime where blueshifted HHG is observed, the optical response undergoes a qualitative transformation. At intermediate fields (24.5~MV/cm), a persistent suppression of reflectivity emerges at positive time delays, extending well beyond the pump pulse duration (Fig.~\ref{fig:m2}b). At the highest field (43.3~MV/cm), this delayed, incoherent component completely dominates the response, and the coherent feature at time zero is hardly visible (Fig.~\ref{fig:m2}c). This transition signifies that the intense MIR pulse irreversibly induces an insulator-to-metal transition within the pulse duration, leaving behind a population of mobile carriers that persists to much longer delays and rendering $\Delta R/R$ asymmetric with respect to the temporal origin \cite{wang2017producing}. The generation of this incoherent metallic state thus strongly suggests a connection between the giant harmonic blueshift and the rapid breakdown of the Mott-insulating state.

We then quantify the field dependence of this persistent metallic response to identify the underlying photo-excitation mechanism (Fig.~\ref{fig:m2}d). The reflectivity change at 1.8~eV, acquired at a delay of 250~fs well beyond the pump-probe overlap, serves as a measure of the long-lived carriers that remain after the drive. At low to moderate fields, the signal amplitude closely follows a standard Zener tunneling formula for carrier generation ($n\propto\exp{(-E_\mathrm{th}/E)}$, dashed line) \cite{keldysh1964ionization,andrei2025subgap}. The competition between the instantaneous coherent response and the delayed incoherent carrier response is further detailed in temporal cuts at 1.8~eV (Fig.~\ref{fig:m2}e) and 2.15~eV (Fig.~\ref{fig:m2}f). At 1.8~eV, $\Delta R/R$ minimum shifts from time zero at low fields (purple shade in Fig.~\ref{fig:m2}e) to progressively later positive delays as the field increases (Fig.~\ref{fig:m2}g), reflecting the growing dominance of absorption (blue shade in Fig.~\ref{fig:m2}e). This is corroborated by the data at 2.15~eV, where the peak coherent enhancement (red shade in Fig.~\ref{fig:m2}f) shifts to earlier negative delays as the incoherent suppression (blue shade in Fig.~\ref{fig:m2}f) increasingly dominates the signal (Fig.~\ref{fig:m2}g). Together, these measurements imply a tunneling-driven electronic state reorganization underlying the HHG blueshift at high driving fields.

\section*{Wavelength-controlled crossover}

To further examine the consequences of the electronic transition on the harmonic spectra, we mapped the frequency shifts across a series of MIR drive wavelengths (Fig.~\ref{fig:m3}; raw emission spectra are shown in Fig.~\ref{sfig:raw hhg spectra}). At 6885~nm, the HHs remain blueshifted above a threshold field, closely mirroring the 7550~nm emission, albeit with a smaller amplitude (Fig.~\ref{fig:m3}a). Shortening the drive to 6200~nm reveals a striking crossover: the harmonics initially redshift at low to moderate fields, then recover and swing through integer positions into a pronounced blueshift as the field grows, with the lower orders persisting in the redshift regime over a wider field range than the higher orders (Fig.~\ref{fig:m3}b). At 5625~nm, the blueshift completely disappears and the spectrum becomes redshift-dominated across the entire accessible field range (Fig.~\ref{fig:m3}c), reproducing the previously documented phenomenology \cite{ikeda2025high}. A multiphoton resonant excitation condition is excluded through Floquet analysis of a three-level model incorporating the excitonic states in Sr$_2$CuO$_3$ \cite{kishida2000gigantic} (Supplementary Information section~\ref{sec:three-level model}), which also fails to reproduce the harmonic frequency shifts in the optical Bloch equations framework (Supplementary Information section~\ref{sec:exciton hhg}). The systematic evolution from pure blueshift at 7550~nm to pure redshift at 5625~nm thus reveals a wavelength- and field-controlled competition between two distinct mechanisms, prompting a deeper investigation into their microscopic origins.

\section*{Many-body nonadiabatic gating}

\begin{figure*}[!htbp]
\centering
\includegraphics{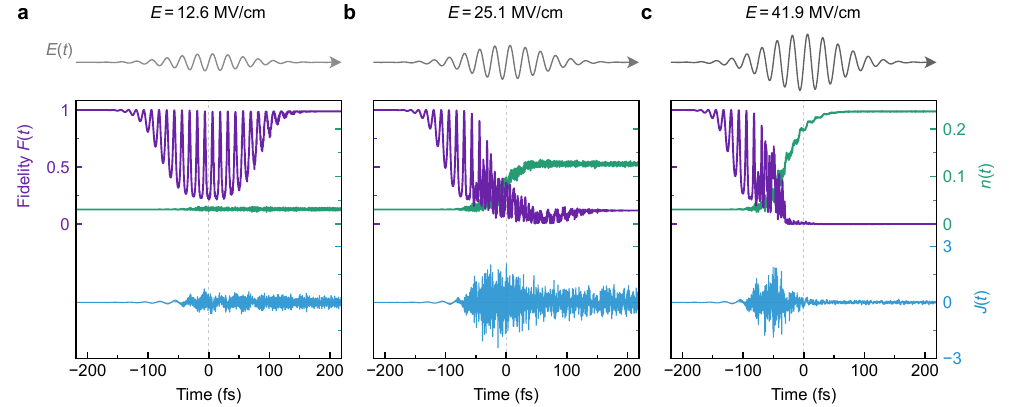}
\caption{\label{fig:m4} Field-dependent many-body dynamics in the driven extended Hubbard chain. Time-dependent exact diagonalization simulation results for peak electric fields of (a), 12.6; (b), 25.1; and (c), 41.9~MV/cm. Upper traces show the driving electric field on a common amplitude scale. Below, synchronized ground-state fidelity \(F(t)\) (purple), doublon density per site \(n(t)\) (green), and the current \(J(t)\) (blue) are displayed. Vertical dashed lines mark the pulse center.}
\end{figure*}

To place the microscopic origin of these shifts on firmer theoretical grounds, we performed time-dependent exact-diagonalization simulations of a driven one-dimensional extended Hubbard model. This framework is known to capture the strong electron correlations in Sr$_2$CuO$_3$ \cite{chen2021anomalously,baykusheva2026quantum}, giving access to the real-time evolution of the many-body state and the associated microscopic current under intense MIR fields. By simulating the system's response to 7550~nm pulses with 100~fs duration across a wide range of driving amplitudes, we directly follow how rapid collapse of the Mott-insulating state during the pulse envelope reshapes the sub-cycle emission dynamics.

The calculated time-dependent currents reveal a dramatic, field-induced restructuring of the emission process (Figs.~\ref{fig:m4} and~\ref{sfig:fidelity}). At low field strengths ($E=12.6$~MV/cm), the electric current peaks near the center of the laser pulse, following the rising envelope of the drive (Fig.~\ref{fig:m4}a). However, as the excitation strength increases into the strong-field regime ($E\geq25.1$~MV/cm), the current oscillations become progressively suppressed on the trailing edge of the pulse (Fig.~\ref{fig:m4}b and c). We emphasize that no explicit radiative damping or dephasing channel is imposed in the simulation, which leads to residual oscillations beyond the pulse when the carrier density is low \cite{murakami2021high}, suggesting this suppression of the current arises from purely unitary many-body dynamics. Notably, this process does not uniquely depend on the dimensionality of the Hubbard model (Supplementary Information section \ref{sec:2d hubbard}), and our analysis is in line with previous results obtained from dynamical mean-field theory \cite{ikeda2025high}, the time-dependent Schrödinger equation \cite{silva2018high}, and variational Monte Carlo \cite{orthodoxou2021high}.

As is well established in the context of atomic and molecular gases, HHG within a finite-duration pulse carries an additional nonadiabatic phase relative to the integer comb \cite{lewenstein1995phase,kan1995spectral}, whose sign is set by the relative electric field variation during a half optical cycle. On the rising edge of the pulse, the field at acceleration and recollision exceeds that at creation, and carriers gain extra phase, generating blueshifted emission, whereas on the falling edge, the field variation reverses and the HH redshifts (Supplementary Information section \ref{sec:nonadiabatic phase}). In a standard medium that remains an active emitter throughout the pulse, these two contributions are sampled symmetrically relative to the pulse center and combine to produce peaks at integer multiples of the drive frequency. 

Unlike HHG in weakly correlated semiconductors, in which independent electron-hole pairs radiate within rigid bands and emission persists as long as carriers can be re-accelerated, harmonic emission in Mott insulators is carried by constructive interference of doublon-holon recombination pathways that remain coherent with the ground state. Because strong correlations suppress coherence among the excited carriers themselves \cite{silva2018high}, sustained emission requires the evolving wavefunction $|\psi(t)\rangle$ to retain its overlap with the Mott-insulating state $|\psi_0\rangle$ (Supplementary Information section~\ref{sec:hubbard fidelity}), quantified by ground-state fidelity $F(t)=|\langle\psi_0|\psi(t)\rangle|^2$, and it ceases once this mutual coherence is lost, even if a dense carrier population persists.

\begin{figure}[tbp]
\centering
\includegraphics[width=0.5\textwidth]{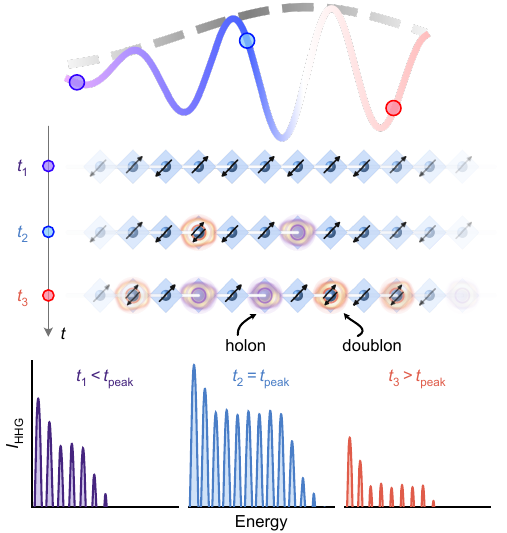}
\caption{\label{fig:m5} Mechanism of harmonic frequency blueshift in a driven one-dimensional Mott insulator. Schematic real-space charge distribution in a Hubbard chain (second row) following long-wavelength MIR excitation (first row; color-coded according to the sign of the harmonic frequency shift) is shown for three instants, as well as the corresponding HHG spectra (third row; $t_\mathrm{peak}$ denotes the peak emission time).}
\end{figure}

In strongly driven Sr$_2$CuO$_3$, this electronic state reconstruction can break the temporal balance between rising- and falling-edge emission, as illustrated in Fig.~\ref{fig:m5}. Our calculations show that the fidelity largely recovers after weak excitation (Fig.~\ref{fig:m4}a) but remains strongly suppressed at high fields (Fig.~\ref{fig:m4}b and c), accompanying the attenuation of trailing-edge current (Fig.~\ref{sfig:subcycle analysis}). These results support a temporal gate that preferentially selects rising-edge emission, favoring a nonadiabatic blueshift that increases with orders (Supplementary Information section~\ref{sec:nonadiabatic solid-state hhg}) and an asymmetric emission profile (Fig.~\ref{sfig:schematic profile}). This mechanism is conceptually analogous to ionization-induced gating in atomic gases \cite{shin1999generation,shin2001nonadiabatic} and highly oriented pyrolytic graphite \cite{chen2026two}, although here the gate closes at a carrier density far below full depletion of the valence band (Supplementary Information section~\ref{sec:carrier density}; Fig.~\ref{fig:m4}). Our analysis further indicates that this behavior is incompatible with a weakly interacting Bloch electron description, pointing to the role of strong electron–electron correlations (Supplementary Information section~\ref{sec:sbe eid}).

Crucially, the nonadiabatic blueshift contrasts sharply with the redshift associated with transient Mott-gap reduction in Sr$_2$CuO$_3$ \cite{ikeda2025high}. Whereas gap renormalization \cite{okamoto2011photoinduced,novelli2014witnessing,matsuzaki2015ultrafast,baldini2020electron} slightly modifies the recombination phase, our opposite and much larger spectral shifts highlight the additional role of many-body coherence loss in gating the emission. The balance between these effects is controlled by the driving wavelength. For a fixed laser pulse duration, longer wavelengths pack fewer optical cycles into the envelope, resulting in a substantially larger variation of the electric field from one half-cycle to the next. This enhances the nonadiabatic phase accumulation on the rising edge of the pulse (Fig.~\ref{sfig:nonadiabatic phase}), allowing the blueshift to dominate over the concurrent redshift. As the wavelength is shortened, the field varies more smoothly from cycle to cycle and the quantum tunneling efficiency increases \cite{andrei2025subgap}, together tipping the balance toward the redshift channel and making it the dominant spectral signature. This dynamic interplay elegantly accounts for the striking wavelength-dependent crossover observed in our experiments. Complementary semiconductor Bloch equations simulations qualitatively reproduce both spectral shifts and their transition when rapid excitation-induced dephasing and carrier-dependent gap narrowing are imposed phenomenologically (Supplementary Information section~\ref{sec:sbe wavelength crossover}).

\section*{Discussion and outlook}
The observation of this macroscopic, nonadiabatic blueshift establishes a new paradigm for understanding and utilizing solid-state HHG. Beyond intensity and polarization, the peak emission energy of HHs serves as a direct, background-free observable of irreversible quantum many-body dynamics. This capability elevates HHG from a probe of single-particle band structure \cite{vampa2015all,lanin2017mapping} and topology \cite{luu2018measurement,heide2022probing,uzan2024observation} to a dynamic, real-time imaging tool for complex collective phenomena rooted in strong electron correlations. More broadly, our work extends the hallmarks of many-body interactions beyond equilibrium thermodynamics by demonstrating a fundamental breakdown of the integer-harmonic rule of nonlinear light emission in the single-particle picture. The union of strong-field spectroscopy and strongly correlated quantum materials thus emerges as a fertile new frontier of condensed matter physics for discovering and controlling novel electronic phases on their fastest timescales. Looking forward, using an intense field to both drive a many-body state far from equilibrium and read out its evolution from the emitted radiation offers a promising route to track ultrafast light-induced phase transitions with unprecedented sub-cycle temporal resolution \cite{borsch2023lightwave,zong2023emerging,heide2024ultrafast}.

\section*{Acknowledgments}
The authors thank G. Vampa, R. Ikeda, H. Okamoto, Y. Murakami, T. Devereaux, R. Noguchi, K. Seyler, and H. Kim for useful discussions. M.Guo acknowledges support from the Eddleman Graduate Fellowship. D.H. acknowledges support from the Institute for Quantum Information and Matter, an NSF Physics Frontiers Center (PHY-1733907). Y.W. acknowledges U.S. Department of Energy, Office of Science, Basic Energy Sciences, under Early Career Award No.~DE-SC0024524. R.S. acknowledges the financial support provided by the Ministry of Science and Technology in Taiwan under project numbers NSTC-114-2124-M-001-009 and NSTC-113-2112-M-001-045-MY3, and Academia Sinica for the budget of AS-iMATE-115-14. The work at the University of Minnesota was funded by the US Department of Energy through the University of Minnesota Center for Quantum Materials, under Grant No. DE-SC-0016371. The simulation used resources of the National Energy Research Scientific Computing Center, a U.S. Department of Energy Office of Science User Facility located at Lawrence Berkeley National Laboratory, operated under Contract No.~DE-AC02-05CH11231.

\section*{Author contributions}
M.Guo performed optical experiments, single-particle simulations, and data analysis. M.Guo and L.W. carried out time-dependent Schrödinger equation simulations supervised by Y.W. M.S., S.K., and R.S. grew and characterized the single crystals of Sr$_2$CuO$_3$. B.Y. and M.Greven synthesized and characterized single crystals of Sr$_2$CuO$_2$Cl$_2$. M.Guo, Y.W., and D.H. wrote the manuscript with input from all other authors. D.H. supervised the project.

\section*{Competing interests}
There are no competing interests to declare.

\section*{Data and materials availability}
The data supporting the findings of this study are provided in the main text or the supplementary materials.

\bibliography{refs}

\clearpage
\onecolumngrid
\raggedbottom

\setcounter{section}{0}
\setcounter{subsection}{0}
\setcounter{figure}{0}
\setcounter{table}{0}
\setcounter{equation}{0}
\renewcommand{\thesection}{\arabic{section}}
\renewcommand{\thesubsection}{\thesection.\arabic{subsection}}
\renewcommand{\thefigure}{S\arabic{figure}}
\renewcommand{\thetable}{S\arabic{table}}
\renewcommand{\theequation}{S\arabic{equation}}

\makeatletter
\@booleantrue\raggedcolumn@sw
\renewcommand{\p@subsection}{}
\let\@hangfrom@section\@hang@from
\def\@hangfroms@section#1#2{#1#2}
\def\l@f@section{\addpenalty{\@secpenalty}\addvspace{.4em plus .1em}}
\makeatother

\begin{center}
{\large\bfseries Supplementary Materials for\\ \manuscripttitle\par}
\vspace{1.5em}
Mingyao~Guo$^{\ast}$,
Luting~Wang,
Murali~Suganya,
Biqiong~Yu,
Martin~Greven,
Sritharan~Krishnamoorthi,
Raman~Sankar,
Yao~Wang,
David~Hsieh$^{\ast}$\\
\small$^\ast$Corresponding author. Email: mguo@caltech.edu; dhsieh@caltech.edu\\
\end{center}

\tableofcontents
\addtocontents{toc}{\protect\SItocShow}

\clearpage

\section{Material growth and characterization}
High-quality Sr$_2$CuO$_3$ single crystals were grown by the optical floating zone method. Initially, the polycrystalline powders were prepared by conventional solid-state reaction using high-purity SrCO$_3$ and CuO in stoichiometric proportions. The hygroscopic SrCO$_3$ was preheated at 1000~$^{\circ}$C for 3~h. The thoroughly mixed powders were calcined in air at 980~$^{\circ}$C, with several intermediate grindings performed to ensure phase purity and compositional homogeneity. The reacted powder was subsequently pressed into cylindrical rods and sintered at 700~$^{\circ}$C to obtain dense and mechanically stable feed rods. Crystal growth was carried out in an optical floating-zone furnace under a flowing oxygen atmosphere, with the feed and seed rods counter-rotated at 10–20~rpm to ensure melt homogeneity. A stable molten zone was translated at a growth rate of 1--3~mm/h, resulting in the formation of a crystalline boule. After growth, the crystal was cooled to room temperature at a controlled rate of 50–100~$^{\circ}$C h$^{-1}$ to minimize thermal stress.

Sr$_2$CuO$_2$Cl$_2$ single crystals were synthesized by standard methods described in \cite{miller1990synthesis}.

\clearpage

\section{Experimental setup}
\label{sec:experimental setup}
\begin{figure*}[hb]
\includegraphics[width=\textwidth]{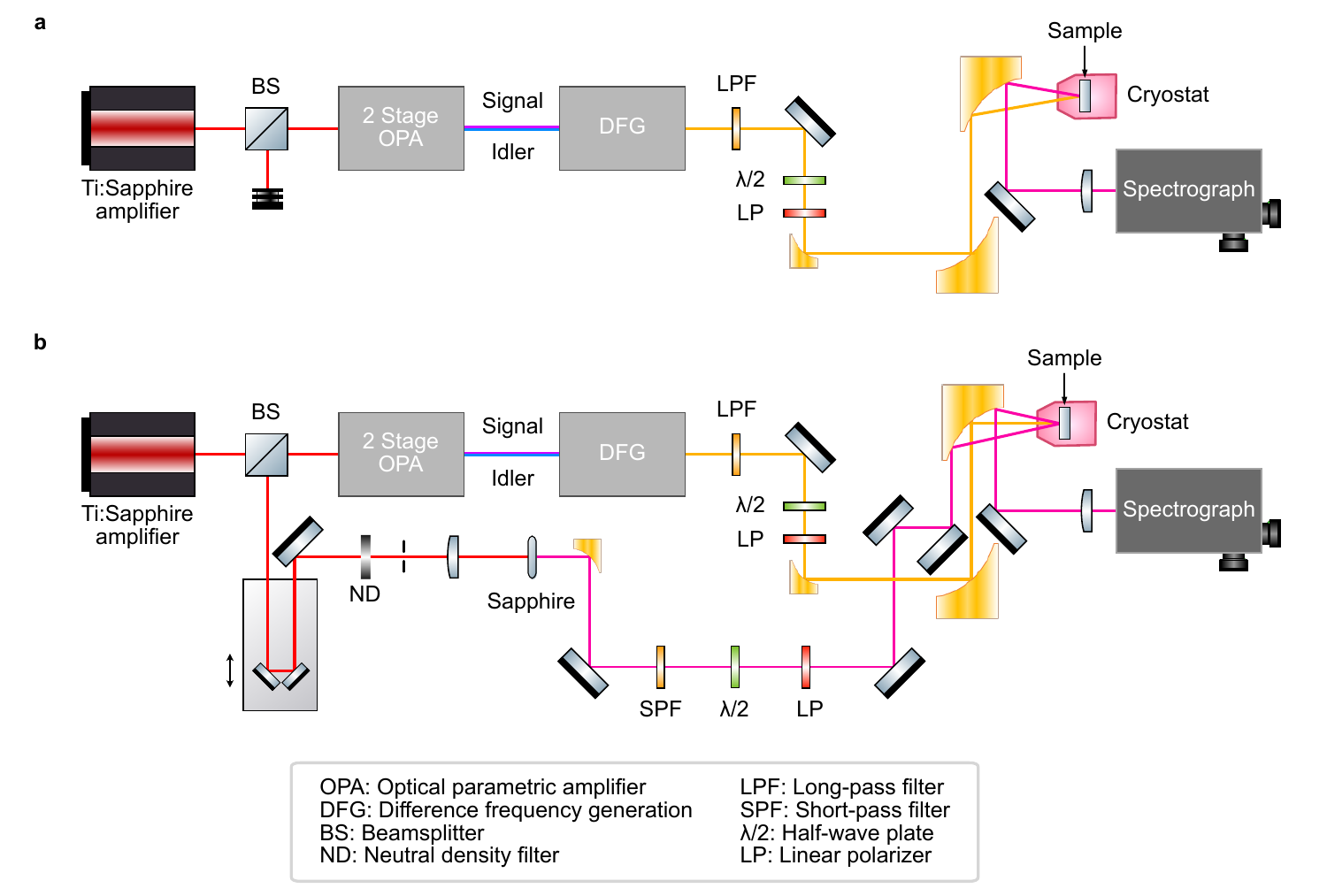}
\caption{\label{sfig:setup} Schematic layouts of the experimental (a) HHG, and (b) transient reflectivity spectroscopy setups.}
\end{figure*}

HHG measurements were carried out using the experimental setup shown in Fig.~\ref{sfig:setup}a at 80~K unless otherwise specified. The setup is based on a femtosecond titanium-sapphire laser amplifier (repetition rate, 1~kHz; pulse duration, 35~fs; center wavelength, 800~nm), which is used to pump a dual-stage optical parametric amplifier (OPA). The output signal/idler pulses are then used to generate MIR pulses by their difference frequency mixing in a GaSe crystal in a noncollinear geometry (pulse duration, $\sim$100~fs). Because signal and idler come from the same OPA without any phase locking, the MIR pulses do not have a stable carrier-envelope phase (CEP). A typical spectrum integrates signals from more than ten thousand pulses, effectively averaging out the CEP-dependent effects. Each IR pulse wavelength was characterized before the measurements with a home-built grating-based spectrometer (Fig.~\ref{sfig:mir spectra}). A germanium long-pass filter was employed to filter out the residual signal and idler pulses. 

HHs were obtained by focusing the IR beams obliquely (incident angle, 10$^{\circ}$) onto the sample by an off-axis parabolic (OAP) mirror. The highest employed vacuum peak driving field of 45~MV/cm is close to the observed long-time damage threshold of Sr$_2$CuO$_3$ and Sr$_2$CuO$_2$Cl$_2$. The resulting reflected spectrum was collected by the same OAP mirror and recorded by a home-built grating-based spectrometer with an indium gallium arsenide detector array (0.8--1.7~$\mu$m) and a cooled silicon charge-coupled device (0.3--1.1~$\mu$m). All spectra are corrected for the reflectance of silver- and aluminum-coated mirrors, transmission of the lens and cryostat window, diffraction efficiency of the grating, and quantum efficiency of the detector.

Transient reflectivity spectroscopy measurements were carried out using the experimental setup shown in Fig.~\ref{sfig:setup}b. The broadband visible pulse was produced via supercontinuum generation seeded by an 800~nm pulse in a 4-mm thick sapphire crystal and filtered by a short-pass filter at 750 nm. The MIR pump and visible probe beams were focused onto the sample using an OAP, and the reflected probe was sent to the same home-built spectrometer for detection. The angle of incidence of the probe is $\sim$10$^{\circ}$, while the pump was set at normal incidence. Both pulses were linearly polarized along the $b$ axis of Sr$_2$CuO$_3$. The temporal chirp dispersion was found to be smooth, fitted by a polynomial function, and corrected through analysis.

\begin{supplementaryfigure}
\includegraphics[width=.75\linewidth]{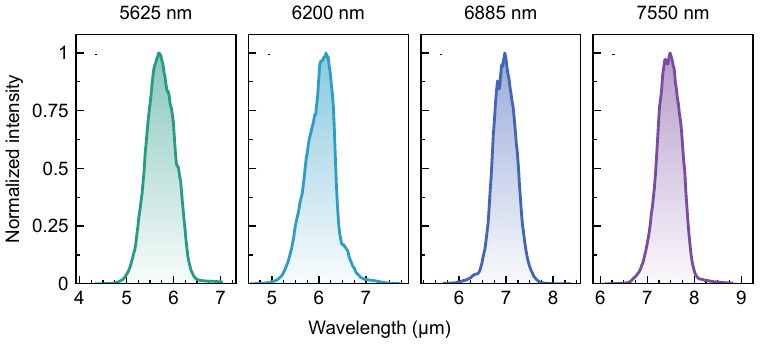}
\caption{\label{sfig:mir spectra} Spectra of the MIR driving pulses. Measured spectra at nominal driving wavelengths used in the study, from left to right, are normalized to their respective peak intensities. The black bar in each panel indicates the spectral resolution of 50~nm.}
\end{supplementaryfigure}

\clearpage

\section{HHG emission profile and determination of central harmonic energy}
\label{sec:HHG profile}
The central emission energy of each harmonic under MIR excitation is extracted from the reflected HHG spectra as follows. Due to the asymmetric non-Gaussian profile of HHG at high field strengths, we use the intensity-weighted first moment (centroid) to define the emission energy
\begin{equation}
\hbar\omega_n^{\mathrm{c}}
   = \frac{\sum_i \hbar\omega_i\, I(\hbar\omega_i)}{\sum_i I(\hbar\omega_i)},
\end{equation}
where the summation runs over the contiguous spectral interval around the local maximum while excluding the spectral background and neighboring orders. This first-moment estimator is robust to the asymmetric profiles of the non-integer harmonics. The resulting peak shift is $\Delta(\hbar\omega_n) = \hbar\omega_n^{\mathrm{c}} - n\hbar\omega_0$. As shown in Fig.~\ref{sfig:hhg profile}, the triangles mark the obtained $\hbar\omega_n^{\mathrm{c}}$ for each spectrum; their progression to higher order with increasing field visualizes the order-dependent blueshift quantified in Fig.~\ref{fig:m1}d.

\begin{supplementaryfigure}
\includegraphics{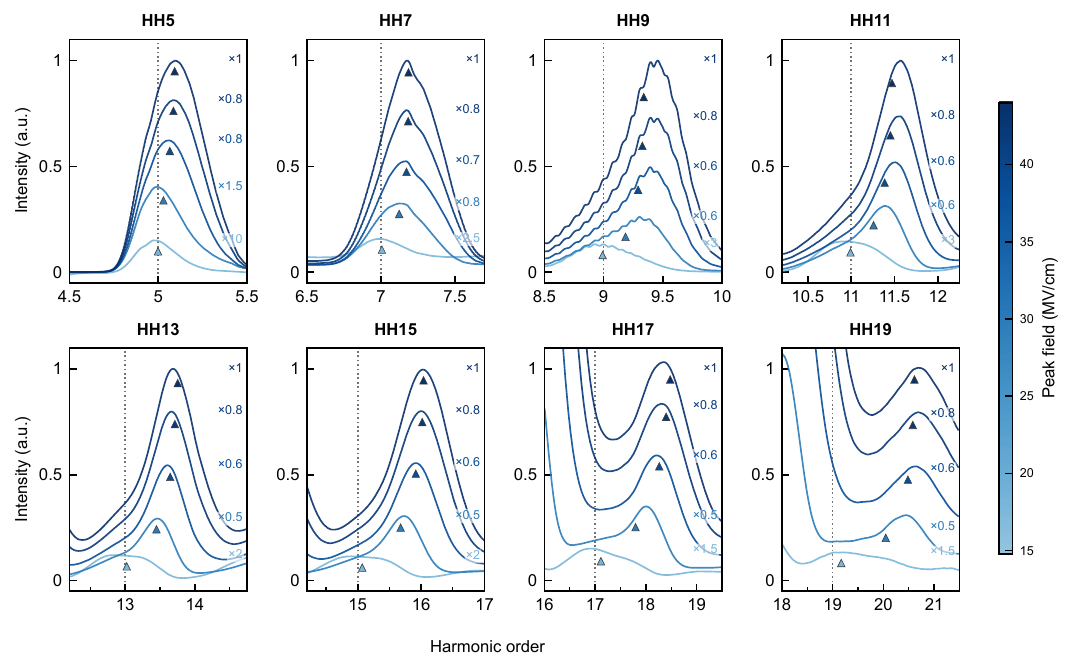}
\caption{\label{sfig:hhg profile} Determination of the harmonic central emission energies and their field dependence with drive wavelength 7550~nm. In each panel the spectra at five peak fields are overlaid. The horizontal axis is the harmonic order $\omega/\omega_0$; grey dotted lines mark the odd integer positions $n$. Triangles beneath each curve mark the central emission energy obtained from the intensity-weighted centroid. With increasing field the central energies shift systematically above the integer positions, by an amount that grows with harmonic order.}
\end{supplementaryfigure}

Beyond quantifying the shift, this asymmetric profile—characterized by a broader tail on the low-energy side—is a direct spectral hint of the nonadiabatic gating mechanism. Figure~\ref{sfig:schematic profile} illustrates how temporal gating towards the rising edge of the pulse naturally dictates this specific line shape by combining the sub-cycle nonadiabatic phase (Sec.~\ref{sec:nonadiabatic phase}) with the electronic state reconstruction (Sec.~\ref{sec:ed}):

\paragraph{Conventional emitters:}
In standard materials that radiate continuously throughout the laser pulse, emission bursts from the rising and trailing edges of the pulse are sampled symmetrically about the pulse center. Consequently, the blueshifted bursts (rising edge) and redshifted bursts (trailing edge) occur in balanced pairs. These equal and opposite nonadiabatic shifts cancel each other out, yielding a symmetric, near-Gaussian spectral line perfectly centered at the integer position $n\hbar\omega_0$ (Fig.~\ref{sfig:schematic profile}a, b).

\paragraph{Driven Mott insulators:}
In Sr$_2$CuO$_3$, the rising optical field drives an ultrafast electronic state reorganization that proliferates doublon-holon pairs and destroys the overlap between the driven many-body state and the Mott-insulating ground state. This truncates the emission and confines it towards the leading half of the pulse (Fig.~\ref{fig:m4} and Sec.~\ref{sec:ed}). This gating strongly suppresses the trailing-edge bursts, leaving only a weak residual emission on the falling edge (Fig.~\ref{sfig:schematic profile}c). The spectral comb is therefore overwhelmingly dominated by the blueshifted (rising-edge) peaks. When these components coherently sum, the resulting line is net-blueshifted and distinctly non-Gaussian, carrying a pronounced low-energy tail composed of the faint, redshifted trailing-edge components (Fig.~\ref{sfig:schematic profile}d). The same temporal gating that produces the blueshift (Fig.~\ref{fig:m5}) thus also imprints the spectral asymmetry, and both are clearly resolved in our measured spectra (Fig.~\ref{sfig:hhg profile}).

\begin{supplementaryfigure}
\includegraphics{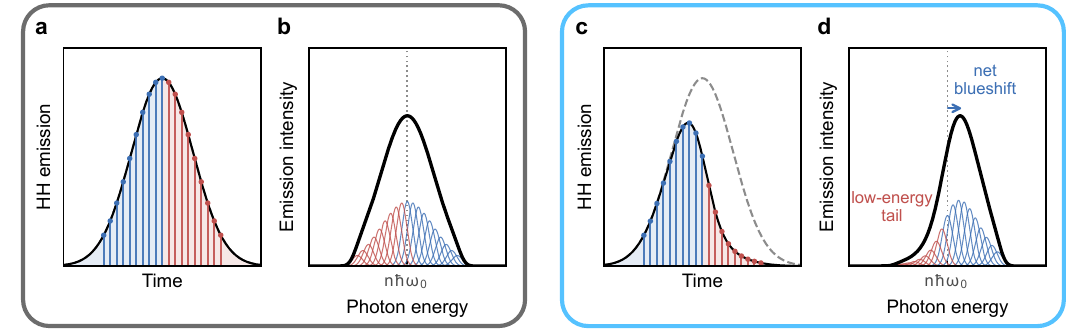}
\caption{\label{sfig:schematic profile} Schematic of the blueshifted, asymmetric (non-Gaussian) harmonic line shape. Each optical half-cycle launches a harmonic burst (stems in a,c); the nonadiabatic phase chirps the emission frequency, so a burst on the rising edge of the pulse is blueshifted ($\omega>n\omega_0$, blue) and one on the trailing edge is redshifted ($\omega<n\omega_0$, red). In the frequency domain each burst is a narrow peak and the spectrum is their sum (black). (a,b) Conventional emitters: the medium radiates over the whole pulse, the blueshifted and redshifted bursts balance, and the comb sums to a symmetric line centered at the integer position $n\hbar\omega_0$. (c,d) Driven Mott insulators: the field-induced electronic state reorganization gates the emission to the rising edge, strongly suppressing the trailing (redshifted) bursts so that only a weak residual remains. The comb is therefore heavily weighted to the blue and sums to a net-blueshifted line that is markedly asymmetric, with a pronounced low-energy tail, reproducing the measured line shapes (Fig.~\ref{fig:m1}a). In (a,c) the dashed grey curve is the pulse envelope and the black curve the (gated) emission; in (b,d) the grey dotted line marks $n\hbar\omega_0$.}
\end{supplementaryfigure}

\clearpage

\section{Benchmark of integer HHG from conventional band semiconductors}
\label{sec:benchmark}
\begin{figure}
\includegraphics{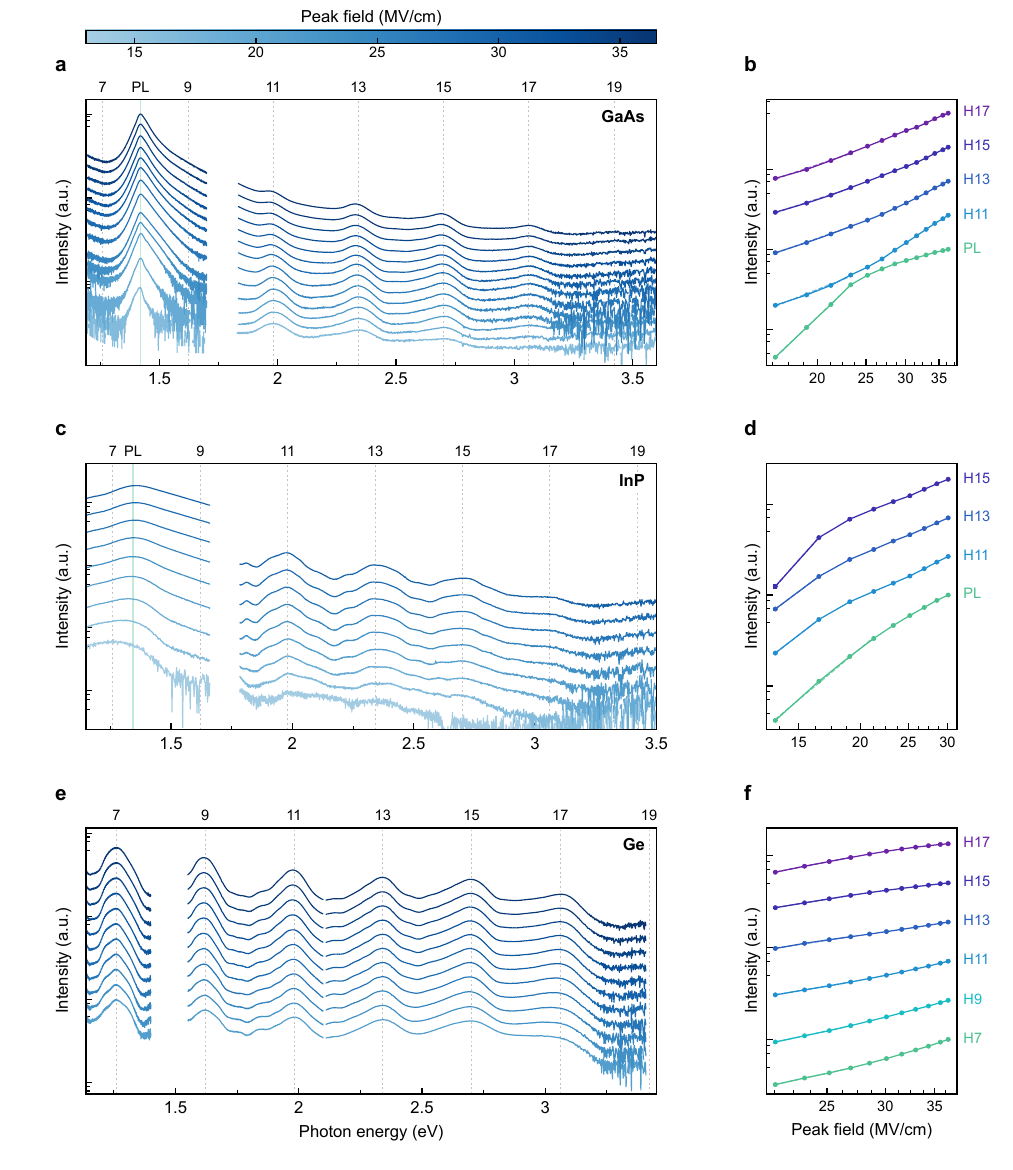}
\caption{\label{sfig:semiconductor benchmark} Field-dependent high-harmonic generation in conventional semiconductors. (a,c,e) Emission spectra of GaAs, InP, and Ge, respectively, under 6885~nm excitation. Color indicates the peak driving field. Spectra are normalized to their maxima within the harmonic range and vertically offset for clarity. Gray dashed lines mark odd harmonic energies, and green lines indicate the nominal PL energies in GaAs and InP. (b,d,f) Corresponding field dependence of the integrated emission in the labeled harmonic bands, together with PL for GaAs and InP (green). Each curve is normalized to its maximum and vertically offset. Intensity axes are logarithmic throughout; the field axes in b, d, and f are also on a log scale. Dashed lines show power-law fits to the low-field points.}
\end{figure}

The optical charge gap of Sr$_2$CuO$_3$ \cite{kim2008bound} and Sr$_2$CuO$_2$Cl$_2$ \cite{zibold1996optical} is approximately 1.5~eV. We therefore first benchmarked its HHG response against GaAs(001) and InP(100), two weakly correlated semiconductors with comparable direct band gaps of approximately 1.42 and 1.34 eV, respectively \cite{beaudoin1997optical}. Under 6885~nm excitation that is identical to the condition in Fig.~\ref{fig:m3}a, both materials exhibit harmonic peaks at odd-integer multiples of the driving frequency throughout the investigated field ranges (Fig.~\ref{sfig:semiconductor benchmark}a,c). Their harmonic intensities increase with the driving field, accompanied by increasing band-edge photoluminescence (PL), which confirms carrier generation. This increasing emission occurs without the pronounced, systematic displacement from integer harmonic positions observed in Sr$_2$CuO$_3$, consistent with earlier reports on GaAs in reflection geometry \cite{xia2021high,sekiguchi2022polarization,sekiguchi2023enhancing}.

Since the proposed many-body temporal gating in this work is associated with carrier excitation and the resulting loss of coherence with the ground state, an important control is to further increase carrier density. We therefore extended the comparison to Ge(100), whose direct optical gap of approximately 0.8~eV \cite{zwerdling1959exciton} provides a substantially lower interband excitation threshold than those of GaAs, InP, Sr$_2$CuO$_3$, and Sr$_2$CuO$_2$Cl$_2$. Ge likewise exhibits integer-order harmonic peaks with increasing driving field (Fig.~\ref{sfig:semiconductor benchmark}e). Thus, reducing the excitation gap does not produce the large non-integer spectral shifts characteristic of Mott-insulating cuprates.

The field-dependent harmonic yields further support a nonperturbative origin of these integer-order spectra (Fig.~\ref{sfig:semiconductor benchmark}b,d,f). In all three semiconductors, power-law fits to the lower-field portion of the measured datasets yield exponents substantially smaller than $2n$ for the $n$-th order, indicating pronounced departures from perturbative scaling. The persistence of integer-order harmonic peaks therefore extends into the nonperturbative regime and cannot be attributed simply to the semiconductor measurements remaining within the weak-field perturbative limit.

Together, these control measurements establish integer HHG in weakly correlated semiconductor benchmarks spanning both comparable and smaller bandgaps. They demonstrate the workability of our optical setup and show that carrier generation and increasing nonlinear emission do not necessarily produce a large harmonic blueshift, supporting the role of correlation-driven electronic dynamics in the nonadiabatic gating proposed for Sr$_2$CuO$_3$.

\clearpage

\section{Non-integer HHG from \texorpdfstring{Sr$_2$CuO$_2$Cl$_2$}{Sr2CuO2Cl2}}
We measured reflected HHG from the quasi-two-dimensional Mott insulator Sr$_2$CuO$_2$Cl$_2$ using 7700~nm MIR pulses polarized along the Cu–O bond direction (Fig.~\ref{sfig:Sr2CuO2Cl2_HHG}a). The harmonic emission exhibits field- and order-dependent blueshifts from the expected odd-integer energies (Fig.~\ref{sfig:Sr2CuO2Cl2_HHG}b), together with non-perturbative scaling of the integrated harmonic intensities (Fig.~\ref{sfig:Sr2CuO2Cl2_HHG}c). These observations extend the blueshifted non-integer HHG found in the quasi-one-dimensional chain compound Sr$_2$CuO$_3$ to a second cuprate Mott insulator.

\begin{supplementaryfigure}
\includegraphics[width=0.5\linewidth]{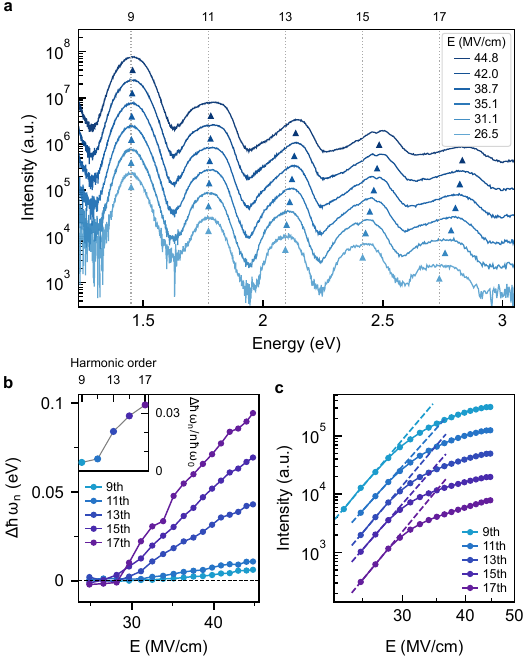}
\caption{\label{sfig:Sr2CuO2Cl2_HHG} Blueshifted non-integer HHG from bulk Sr$_2$CuO$_2$Cl$_2$. (a) Reflected HH emission spectra at various peak field strengths driven by 7700~nm MIR pulses polarized along the Cu-O bond direction, vertically offset for clarity. The grey dotted lines denote the expected emission energies of odd-order HHs, and the triangular markers denote the measured harmonic energies. (b) HH central emission energy shifts $\Delta\hbar\omega_n$ as a function of peak electric field, referenced to their expected energies. The inset displays the order-dependent relative peak shifts $\Delta\hbar\omega_n/n\hbar\omega_0$ at the maximum field. The order-dependent blueshift mimics that observed in Sr$_2$CuO$_3$ (Fig.~\ref{fig:m1}), corroborating the intrinsic and generic nature of the frequency shift in Mott insulators. (c) Integrated HH intensity as a function of peak electric field, vertically offset for clarity. The dashed lines are power-law fits to signals at low field strengths ($\lesssim$28~MV/cm), giving exponents $\alpha_9=5.1$, $\alpha_{11}=4.8$, $\alpha_{13}=5.1$, $\alpha_{15}=5.0$, and $\alpha_{17}=5.2$ in $I_n\propto E^{2\alpha_n}$.}
\end{supplementaryfigure}

\clearpage

\section{Wavelength-dependent non-integer HHG}
We measured field-dependent HHG spectra of Sr$_2$CuO$_3$ at driving wavelengths of 6885, 6200, and 5625~nm, with the MIR polarization along the Cu–O chains. At 6885~nm, the harmonics blueshift above a threshold field (Fig.~\ref{sfig:raw hhg spectra}a). At 6200~nm, they initially redshift and subsequently cross into the blueshift regime as the field increases, with lower orders remaining redshifted over a wider field range (Fig.~\ref{sfig:raw hhg spectra}b). At 5625~nm, the shifts remain predominantly red throughout the measured range (Fig.~\ref{sfig:raw hhg spectra}c). These spectra underlie the emission-energy shifts summarized in the main text (Fig.~\ref{fig:m3}).

\begin{supplementaryfigure}
\includegraphics{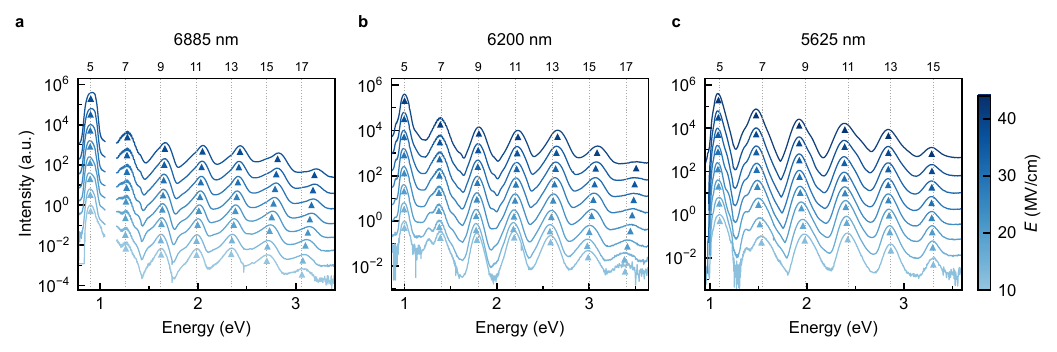}
\caption{\label{sfig:raw hhg spectra} Field-dependent reflected HHG spectra of Sr$_2$CuO$_3$ driven by (a)~6885~nm, (b)~6200~nm, and (c)~5625~nm pulses polarized along the Cu-O chains. In each panel, eight spectra at increasing peak field strength are shown, vertically offset by a constant for clarity. The grey dotted lines denote the expected emission energies of odd-order HHs, and the triangular markers denote the measured harmonic energies. With increasing field, the harmonic peaks evolve away from these integer positions: a pure blueshift above a threshold field at 6885~nm~(a); a redshift at low-to-moderate fields that crosses over into a blueshift at the highest fields for 6200~nm~(b); and a pure redshift at 5625~nm~(c). The peak shifts extracted from these spectra are quantified in Fig.~\ref{fig:m3}.}
\end{supplementaryfigure}

\clearpage

\section{Polarization-dependent HHG}
To establish the directional character of the harmonic emission, we measured the HHG yield as a function of the incident 7550~nm MIR linear polarization, rotated with respect to the crystallographic axes. Figure~\ref{sfig:polarization dependence} shows polar plots of the integrated intensity of HH5, HH9, HH13 and HH17. All four orders display a clear two-fold pattern that sharply peaks when the driving field is polarized along the $b$ axis, i.e.\ along the CuO chains, and is strongly suppressed along the orthogonal $c$ axis. This anisotropy directly reflects the highly quasi-one-dimensional electronic structure of Sr$_2$CuO$_3$: the field-driven intersite currents responsible for HHG are heavily confined to the chain direction, so harmonics are produced efficiently only when the field has a substantial component along $b$. The pattern is consistent across all orders, from the lowest (HH5) to the highest (HH19) shown, indicating that the observed harmonics share the same chain-confined generation mechanism. Accordingly, the driving field is aligned along the $b$ axis throughout the main text to maximize the harmonic yield.

\begin{supplementaryfigure}
\includegraphics{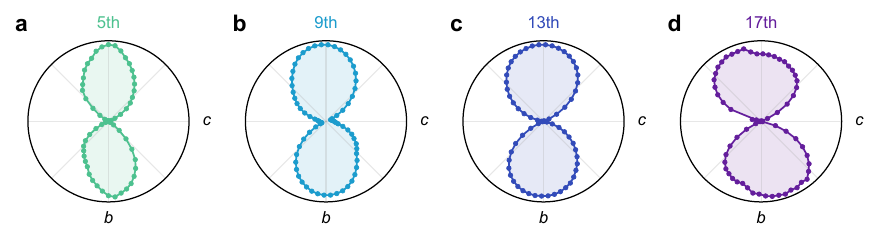}
\caption{\label{sfig:polarization dependence} Polarization dependence of the individual HHs from Sr$_2$CuO$_3$. Polar plots of the integrated emission intensity of (a) HH5, (b) HH9, (c) HH13 and (d) HH17 as a function of the incident 7550~nm MIR linear polarization angle. The radial coordinate is the harmonic intensity (arbitrary units, normalized within each panel) and the angular coordinate is the in-plane polarization direction; $b$ and $c$ denote the crystallographic axes, with the CuO chains running along the $b$ axis. Colors follow the convention of Fig.~\ref{fig:m1}.}
\end{supplementaryfigure}

\clearpage

\section{Non-perturbative scaling of HHG}
Figure~\ref{sfig:field dependence} supplements the field-dependence data presented in Fig.~\ref{fig:m1}c by resolving the distinct scaling behaviors of individual harmonic orders: HH7, HH11, HH15 and HH19. At low driving field, the integrated intensity of each harmonic grows according to a power-law relationship, $I_n\propto E^{2\alpha_n}$ (dashed lines), before exhibiting a pronounced saturation behavior at high fields where the measured signals fall below the power-law extrapolation. The fitted scaling exponents, $\alpha_{7}=3.9$, $\alpha_{11}=5.2$, $\alpha_{15}=5.5$ and $\alpha_{19}=5.4$, are all far below the value $\alpha_n=n$ required for an $n$th-order perturbative process. Together with the low-field exponents obtained for HH5, HH9, HH13 and HH17 in Fig.~\ref{fig:m1}c, these results unequivocally demonstrate that the HH emission is generated in a strongly non-perturbative regime across the entire observed spectral bandwidth.

\begin{supplementaryfigure}
\includegraphics{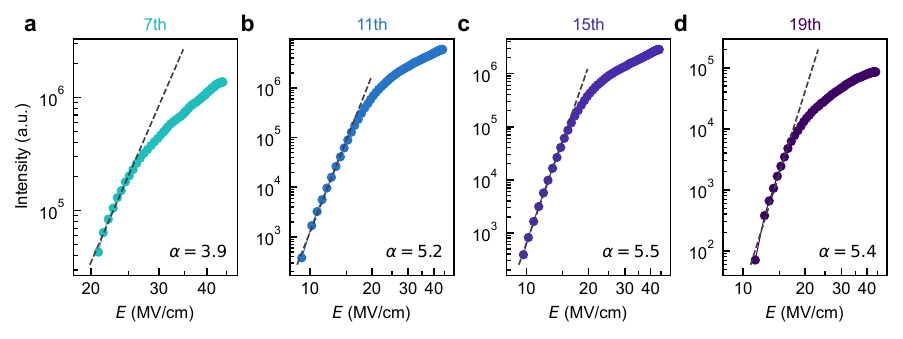}
\caption{\label{sfig:field dependence} Field scaling of the individual HHs in Sr$_2$CuO$_3$. Integrated intensity of (a) HH7, (b) HH11, (c) HH15 and (d) HH19 as a function of the peak driving field $E$ (log-log). Dashed lines are power-law fits to the low-field data, $I_n\propto E^{2\alpha_n}$, with the fitted exponent $\alpha_n$ indicated in each panel. Colors follow the convention of Fig.~\ref{fig:m1}.}
\end{supplementaryfigure}

\clearpage

\section{Temperature-dependent HHG emission energy}
\label{sec:HHG T dep}
To exclude a thermal origin for the observed blueshift, we investigated the temperature dependence of the harmonic emission energies. If the field-induced blueshift were merely a consequence of transient laser heating, the harmonic energies would then exhibit a strong dependence on the lattice temperature. To test this, we continuously varied the base temperature at fixed driving conditions and searched for a comparable shift.

Figure~\ref{sfig:t dep}a displays the resulting central energies of HHG under 7700~nm driving from Sr$_2$CuO$_3$ from 80~K up to 320~K, plotted as the relative shift with respect to the temperature-averaged value of each order. Across the entire range, none of the harmonics displays a systematic temperature-dependent shift. Instead, the peak energies merely fluctuate about their mean values by no more than a few meV, with no monotonic trend (the weak HH19 shows a somewhat larger scatter owing to its low signal-to-noise ratio). This residual scatter is consistent with run-to-run noise from external fluctuations rather than an intrinsic temperature dependence, and its magnitude is more than an order of magnitude smaller than the blueshift measured in the field-dependent experiments (Fig.~\ref{fig:m1}d). We observe a similarly temperature-insensitive behavior from Sr$_2$CuO$_2$Cl$_2$ (Fig.~\ref{sfig:t dep}b), where HH energies also fluctuate by only a few meV with no systematic trend. This robust thermal stability therefore confirms that the strongly blueshifted emission is an intrinsic consequence of the field-driven many-body charge dynamics, rather than a trivial macroscopic heating effect.

\begin{supplementaryfigure}
\includegraphics{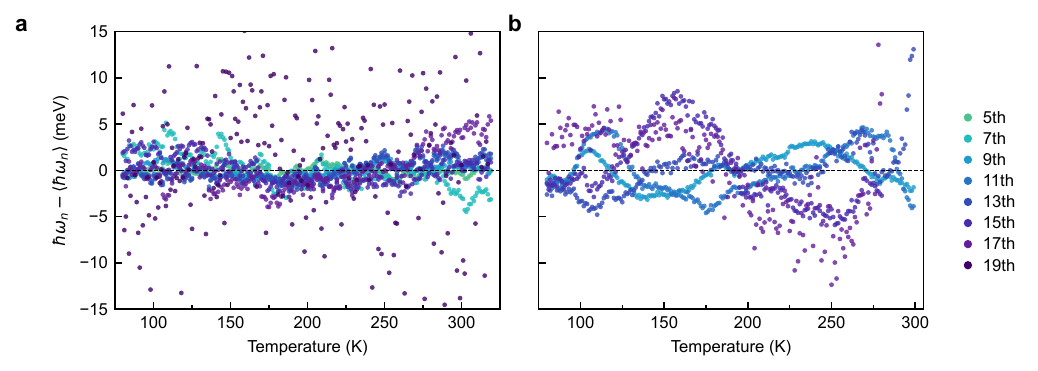}
\caption{\label{sfig:t dep} Thermal stability of harmonic emission energies under fixed MIR driving conditions identical to Figs.~\ref{fig:m1} and ~\ref{sfig:Sr2CuO2Cl2_HHG} at 42~MV/cm. (a) Temperature dependence of the relative harmonic peak energy shift $\hbar\omega_n-\langle\hbar\omega_n\rangle$ from Sr$_2$CuO$_3$ measured from 80 to 320~K. (b) Same as (a) but from Sr$_2$CuO$_2$Cl$_2$.}
\end{supplementaryfigure}

\clearpage

\section{Broadband transient differential reflectivity spectra}
\label{sec:trR}
Figure~\ref{sfig:full trR} presents the complete driving-field dependence of the transient differential reflectivity spectra, complementing the three representative maps of Fig.~\ref{fig:m2}a–c in the main text. Across this series, the optical response evolves continuously from a coherent to an incoherent character as the electric field increases. At the lowest fields, the signal is essentially confined to the pump–probe temporal overlap and leaves no long-lived residual, reflecting an instantaneous, non-dissipative distortion of the Mott state. As the field is raised into the regime where blueshifted harmonics emerge, a persistent change develops at positive delays that extends well beyond the pulse duration. At the highest fields, this delayed, incoherent component dominates the response, while the coherent time-zero feature is nearly extinguished.

This continuous evolution provides a direct, real-time view of the tunneling-driven electronic state reconstruction that underlies the nonadiabatic blueshift. The growth of the long-lived reflectivity change marks the irreversible accumulation of photo-excited doublon–holon pairs that bleach the Mott-gap absorption and metallize the insulating ground state. Because this carrier population is generated within the pulse, it destroys the ground-state coherence required for sustained HH emission, thereby gating the radiation towards the rising edge of the driving field and producing the order-dependent blueshift. The field range over which this persistent component sets in is comparable to the onset of the harmonic blueshift (Fig.~\ref{fig:m1}d), corroborating that both phenomena originate from the same field-induced electronic transition. Furthermore, the gradual field dependence observed here mirrors the continuous progression of current and carrier dynamics captured by our exact diagonalization simulations (Sec.~\ref{sec:ed}).

To further support the validity of the tunneling picture, we estimate the Keldysh parameter $\gamma_K$ under our experimental conditions. In a Mott insulator, this adiabaticity parameter is defined as $\gamma_K=\hbar\omega/eE\xi$ \cite{oka2012nonlinear,andrei2025subgap}, where $\omega$ is the pump angular frequency, $E$ is the pump electric field, $e$ is the elementary charge, and $\xi$ is the doublon-holon correlation length. Analytical theory for the one-dimensional Hubbard model yields $\xi=\ln^{-1}{(U/4.377t_h)}a$ ($U$ is the on-site Coulomb interaction strength, $t_h$ is the electron hopping amplitude, and $a$ is the distance between nearest neighbors) \cite{stafford1993scaling}. Using $U/t_h=8$ for Sr$_2$CuO$_3$, we obtain $\xi=1.658a$. Because the excitation photon energy $\hbar\omega=0.164$~eV lies far below the charge gap, the imaginary part of the dielectric function is negligible ($\epsilon_2\approx0$). Based on the published equilibrium reflectivity $R=0.2$ at this photon energy \cite{baykusheva2026quantum}, we calculate the internal penetrated electric field to be $E_\mathrm{in}=0.55E_\mathrm{vacuum}$. Figure~\ref{fig:m2}d indicates a considerable onset of carrier density at a vacuum field of $\sim E=16$~MV/cm. At this field strength, the Keldysh parameter $\gamma_K=0.287$ is far smaller than unity, locating the light-matter interaction within the deep quantum tunneling regime \cite{kane1960zener,keldysh1964ionization,oka2012nonlinear,andrei2025subgap}, consistent with the Zener-like carrier generation observed in our data.

\begin{supplementaryfigure}
\includegraphics{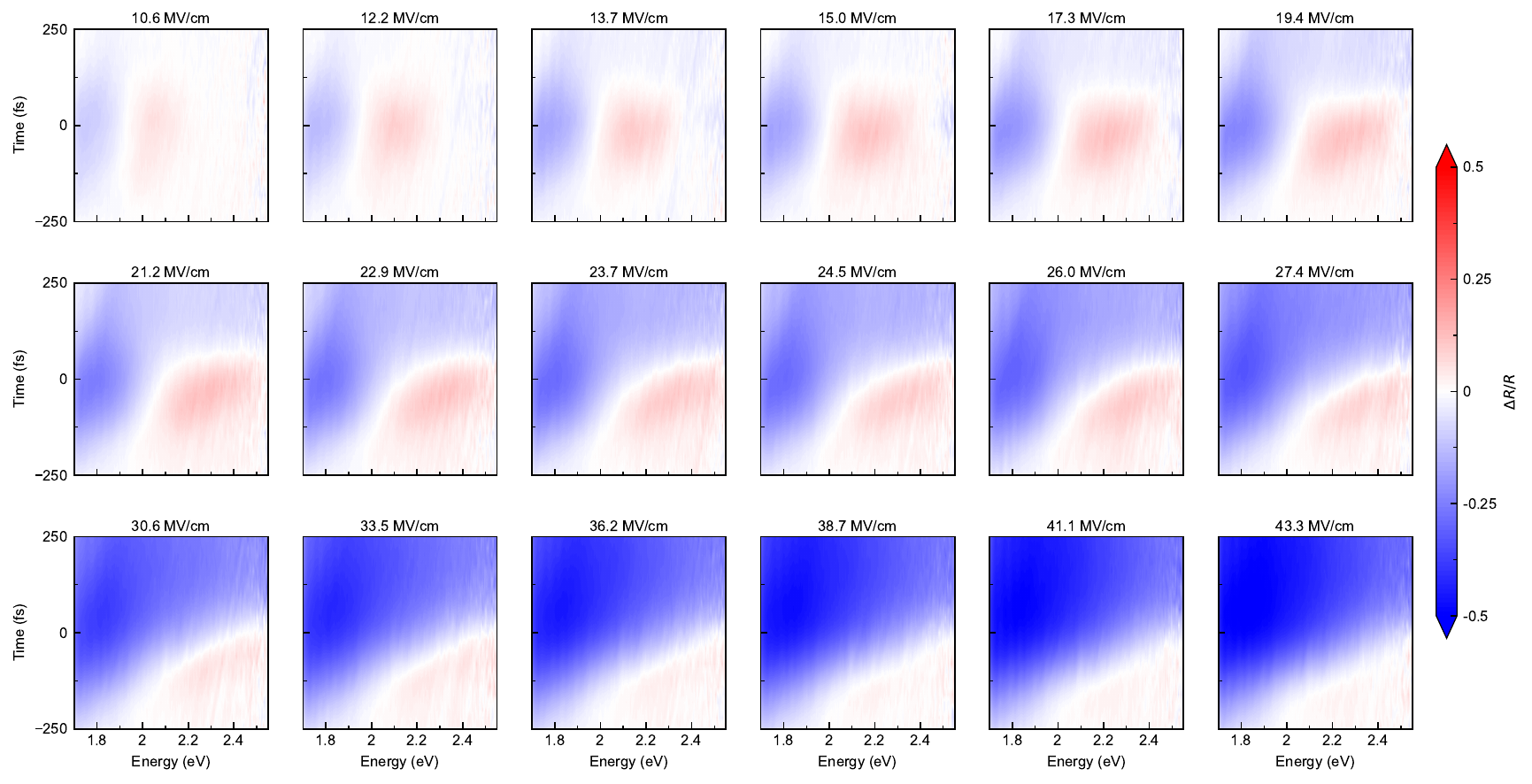}
\caption{\label{sfig:full trR} Complete driving-field series of broadband transient differential reflectivity spectra. Maps of $\Delta R/R$ (probe photon energy vs.\ pump–probe delay) for all measured peak driving fields, from 10.6 to 43.3~MV/cm (labeled on each panel), plotted on the same color scale as Fig.~\ref{fig:m2}a–c.}
\end{supplementaryfigure}

\clearpage

\section{Estimation of carrier density}
\label{sec:carrier density}
\begin{figure}[!hb]
\includegraphics[width=5in]{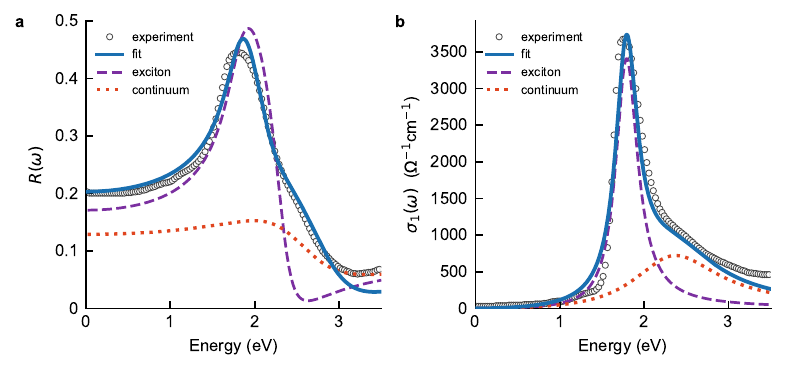}
\caption{\label{sfig:equilibrium R} Equilibrium optical properties of Sr$_2$CuO$_3$. (a) Near-normal-incidence reflectivity $R(\omega)$ and (b) optical conductivity $\sigma_1(\omega)$ of Sr$_2$CuO$_3$ (gray circles, digitized from Baykusheva et al. \cite{baykusheva2026quantum}) fitted with a two-oscillator Lorentzian model (blue). The total response is decomposed into the Hubbard exciton bound below the gap edge (purple dashed line, $E_1=1.79$~eV) and the upper-Hubbard-band continuum (red dotted line, $E_2=2.37$~eV).}
\end{figure}

To definitively rule out macroscopic ground-state depletion—the canonical temporal gating mechanism in atomic gases \cite{shin1999generation,shin2001nonadiabatic} and highly oriented pyrolytic graphite \cite{chen2026two}—as the driver for the observed blueshift, we quantified the persistent carrier density generated during the HHG process. To this end, we first establish a quantitative baseline for the equilibrium optical responses of Sr$_2$CuO$_3$. Both the reflectivity $R(\omega)$ and optical conductivity $\sigma_1(\omega)$ \cite{baykusheva2026quantum} are simultaneously described by a two-oscillator Lorentz model
\begin{equation}
\tilde\sigma(\omega)=\sum_{j=1,2}\frac{-i\varepsilon_0\omega S_j}{E_j^2-(\hbar\omega)^2-i\gamma_j\hbar\omega},
\label{eq:sigma}
\end{equation}
where $\varepsilon_0$ is the vacuum permittivity, and each oscillator $j$ is specified by its resonance energy $E_j$, its strength $S_j$, and its linewidth $\gamma_j$. The spectra are dominated by a single asymmetric absorption peaking near $1.8$~eV, and this line shape is consistent with other optical studies of Sr$_2$CuO$_3$ \cite{kim2008bound,ikeda2025high}. A single parameter set reproduces both quantities across the measured range (Fig.~\ref{sfig:equilibrium R}): a sharp low-energy oscillator at $E_1=1.79$~eV ($\gamma_1=0.31$~eV) corresponding to the bound Hubbard exciton and a broad higher-energy oscillator at $E_2=2.37$~eV ($\gamma_2=1.23$~eV) characteristic of a doublon-holon excitation continuum across the Mott gap; these assignments and parameter values from the fitting are also in quantitative agreement with a recent work \cite{ikeda2025high}.

\begin{figure}
\includegraphics[width=\textwidth]{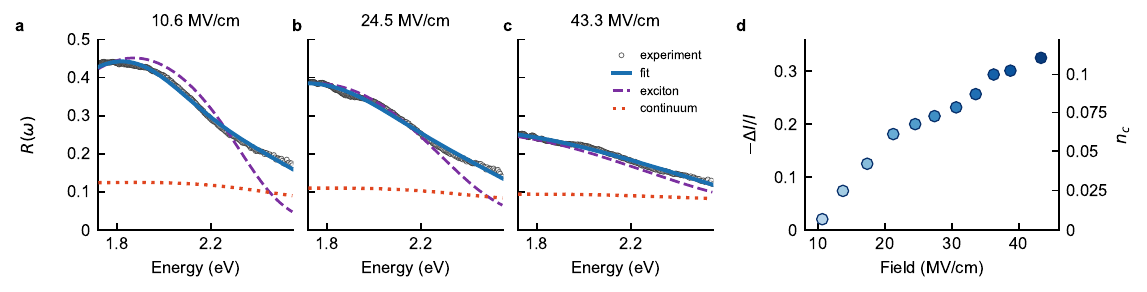}
\caption{\label{sfig:carrier density} Estimation of the carrier density generated in strongly driven Sr$_2$CuO$_3$. (a–c) Transient reflectivity $R(\omega)$ at a pump-probe delay of 250~fs (gray circles), obtained by referencing the measured $\Delta R/R$ to the equilibrium $R(\omega)$ of Fig.~\ref{sfig:equilibrium R}, for peak driving fields of (a) 10.6, (b) 24.5, and (c) 43.3~MV/cm. Solid blue lines are the two-Lorentzian fit over the probe window. (d) Fractional loss of integrated spectral weight of the two Lorentzian oscillators as a function of the peak driving field. The right axis displays the corresponding photo-induced carrier density per Cu site.}
\end{figure}

We then tracked the evolution of this response under photoexcitation at the same MIR driving fields used for HHG. To isolate the incoherent carrier population from instantaneous strong-field effects and minimize thermal artifacts, we analyzed the transient reflectivity at a pump-probe delay of $250$~fs \cite{matsuzaki2015ultrafast,ikeda2025high}. Fitting these transient spectra with the same two-oscillator model while fixing the continuum position and width to their equilibrium values to avoid parameter degeneracy in the narrow window reveals a monotonic bleaching and broadening of the exciton resonance, alongside a suppression of the continuum spectral weight with increasing field strength (Fig.~\ref{sfig:carrier density}a-c). This spectral weight transfer out of the probe window into the low-energy regime is a hallmark of the photo-induced insulator-metal transition. 

With the fitted parameters from $R(\omega)$, we then calculated the corresponding $\sigma_1(\omega)$ and extracted the fractional loss of the integrated spectral weight of the two Lorentzian oscillators (Fig.~\ref{sfig:carrier density}d). In a one-dimensional single-band Hubbard model, the sum rule for the upper Hubbard band gives its spectral weight of the order of $t_h^2/U$ \cite{eskes1994two}
\begin{equation}
w=4\ln2\frac{t_h^2}{U}(n^2+n\frac{\sin2\pi n}{2\pi}-2\frac{\sin^2\pi n}{\pi^2})
\label{eq:sw}
\end{equation}
where $n$ is the occupancy per site ($n=1$ corresponds to half-filling). By numerically inverting this relation (Eq.~\eqref{eq:sw}), we converted the measured weight loss $-\Delta I/I$ into the carrier density $n_c$ apart from half-filling (right axis of Fig.~\ref{sfig:carrier density}d). Crucially, even at the highest field applied in our HHG experiment ($43.3$~MV/cm), the generated carrier density $n_c$ only reaches $\sim0.11$ carriers per Cu site. The valence band population thus remains more than $\sim$90\% intact throughout the pulse, far from the depletion required to truncate the harmonic emission \cite{chen2026two}. Therefore, we exclude the scenario in which the state blocking alone directly gates the HHG and leads to the observed blueshift in emission energy.

\clearpage

\section{Propagation effects on harmonic spectral shifts}
\label{sec:propagation}
The optical response of Sr$_2$CuO$_3$ changes during strong-field excitation, potentially modifying the phase of both the MIR driving field and the emitted harmonics. These time-dependent phases can shift the detected harmonic energies through effects associated with self- and cross-phase modulation. To estimate their magnitude, we first reconstructed the transient complex refractive index during the MIR excitation from the broadband differential reflectivity spectra. We then calculated the phase accumulated through MIR coupling into the sample, harmonic outcoupling, and coherent emission from different depths \cite{kolesik2024propagation,kolesik2025reflection}, and obtain the corresponding instantaneous energy shifts from its time derivative. Finally, we complemented the conclusion drawn from the calculation by a comparison of the reflected MIR spectra at different driving fields.

\subsection{Reconstruction of transient optical constants}
The two-oscillator model introduced in Sec.~\ref{sec:carrier density} describes the bright Hubbard exciton and the doublon–holon continuum. To account for additional spectral changes during the pump–probe overlap, we extended this model by including the first-order Floquet sidebands of the optically dark, even-parity exciton:
\begin{equation}
    \varepsilon(\omega,t)=\varepsilon_\infty+
\sum_{j=\mathrm{exc},\mathrm{cont},-,+}
\frac{S_j(t)}
{E_j^2(t)-(\hbar\omega)^2-i\gamma_j(t)\hbar\omega}.
\end{equation}
Here, $S_j$, $E_j$, and $\gamma_j$ denote the oscillator strength, resonance energy, and linewidth, respectively. We fixed $\varepsilon_\infty=3.322$ and used the exciton and continuum resonance energies obtained from fitting to equilibrium spectra above. The sideband energies were constrained by the even-parity exciton energy $E_{\mathrm{even}}=1.81$~eV and the driving photon energy $\hbar\Omega=0.1642$~eV, placing the two sidebands at approximately 1.646 and 1.974~eV. We imposed equal sideband oscillator strengths following a Gaussian envelope with intensity FWHM of 100~fs, while the amplitude and a common sideband linewidth were fitted separately for each driving field.

\begin{figure*}[ht]
\includegraphics{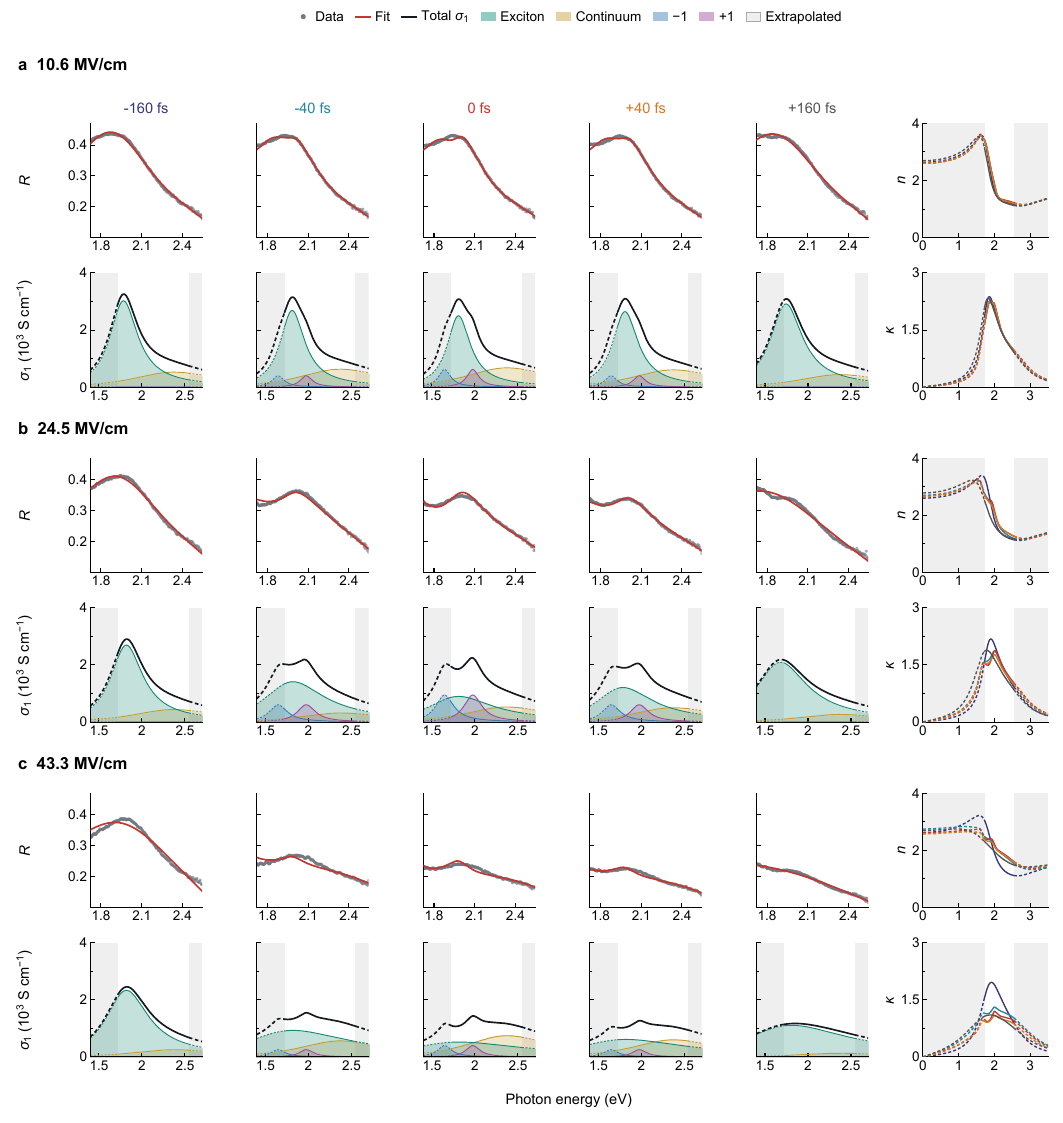}
\caption{\label{sfig:reconstructed optical constants} Reconstructed transient optical constants. (a-c) Spectra of reflectivity, real-part of optical conductivity, real and imaginary parts of refractive index for MIR fields of 10.6, 24.5, and 43.3~MV/cm, respectively. In each group, the first five columns correspond to delays of $-160$, $-40$, 0, 40, and 160~fs. The upper row compares the absolute reflectivity (gray points) with the fitted model (red curves). The lower row shows the reconstructed real optical conductivity $\sigma_1$ (black), with shaded contributions from the bright exciton (green), continuum (ochre), and $-1$ and $+1$ Floquet sidebands of the dark exciton (blue and purple). The rightmost column shows the reconstructed refractive index $n$ (upper row) and extinction coefficient $\kappa$ (lower row); curve colors match the delay headings. Gray backgrounds and dashed curves identify model extrapolation beyond the reflectivity fitting window.}
\end{figure*}

Figure~\ref{sfig:reconstructed optical constants} presents the results for three representative driving fields of 10.6, 24.5, and 43.3~MV/cm at pump–probe delays of $-160$, $-40$, $0$, $40$, and $160$~fs. The model reproduces the principal spectral features and their evolution with driving field and delay very well. The corresponding real part of optical conductivity, calculated as $\sigma_1(\omega,t)=\varepsilon_0\omega\,\mathrm{Im}\,\varepsilon(\omega,t)$, is shown together with its decomposition into the bright-exciton, continuum, and two Floquet-sideband contributions. From the fitted dielectric function, we reconstructed the complex refractive index $n(\omega,t)+i\kappa(\omega,t)=\sqrt{\varepsilon(\omega,t)}$. The resulting $n$ and $\kappa$ are shown over photon energies from 0 to 3.5~eV for the same five delays (Fig.~\ref{sfig:reconstructed optical constants} rightmost column). The reconstructed optical constants provide the input for estimating the propagation-induced harmonic energy shifts described below.

\subsection{Propagation-induced emission energy shifts}
We approximate the near-normal reflection geometry by normal incidence and treat the sample as a semi-infinite medium with an excitation profile uniform in depth. This is justified by the fact that the MIR attenuation depth of $8.4$–$41.6~\mu\mathrm{m}$ extrapolated from the reconstructed spectra greatly exceeds the reflection coherence lengths of $47$–$145$~nm for HH5–HH19, resulting in less than $1.6\%$ intensity loss over the coherence length scale. At each delay, the reconstructed optical constants are held fixed during propagation through the emitting region. We prescribe a coherent harmonic source with uniform magnitude in depth, whose phase follows (q) times the local MIR phase, and hold its additional intrinsic emission phase constant. This isolates the phase contribution arising from the changing optical response.

For harmonic order $q$, we define
\begin{equation}
    N_0(t)=N(\Omega,t),\qquad
N_q(t)=N(q\Omega,t),\qquad
t_{\mathrm{MIR}}(t)=\frac{2}{1+N_0(t)}.
\end{equation}
The complex MIR transmission coefficient contributes a phase $q\arg t_{\mathrm{MIR}}$ to the harmonic source. Radiation generated at depth $z$ additionally acquires the factor
\begin{equation}
    \exp\left[
ik_q(n_0+n_q)z-k_q\kappa_qz
\right],
\qquad k_q=\frac{q\Omega}{c}.
\end{equation}
Here, the MIR and harmonic propagation phases add in reflection because the driving field travels into the sample and the detected harmonic radiation returns to the surface. 

The coherent sum over source depths gives
\begin{equation}
    \int_0^\infty e^{-D_qz}dz=\frac{1}{D_q},
\qquad
D_q(t)=k_q\left[\kappa_q(t)-i\bigl(n_0(t)+n_q(t)\bigr)\right].
\end{equation}
Harmonic outcoupling contributes an additional factor $1/(1+N_q)$, obtained by combining the source-radiation factor $1/N_q$ with the Fresnel transmission coefficient $2N_q/(1+N_q)$. Up to real amplitude factors and time-independent constants, the propagation-dependent complex factor is therefore
\begin{equation}
    H_q(t)=
\frac{\exp\left[iq\arg t_{\mathrm{MIR}}(t)\right]}
{\left[1+N_q(t)\right]D_q(t)}.
\end{equation}

With the field convention $e^{-i\omega t}$, the instantaneous energy shift is
\begin{equation}
    \Delta E_q(t)=-\hbar\frac{d}{dt}\arg H_q(t).
\end{equation}
It separates into the three contributions shown in the harmonic-resolved calculation:
\begin{equation}
\Delta E_q^{\mathrm{MIR}}(t)=\hbar q\operatorname{Im}
\frac{\dot N_0}{1+N_0},\
\Delta E_q^{\mathrm{escape}}(t)=\hbar\operatorname{Im}
\frac{\dot N_q}{1+N_q},\
\Delta E_q^{\mathrm{depth}}(t)=\hbar\operatorname{Im}
\frac{\dot D_q}{D_q}.
\end{equation}
These terms describe the changing MIR entry phase, harmonic radiation and escape phase, and coherent depth integral, respectively, with positive (negative) values denoting blueshift (redshift).

\begin{figure*}[ht]
\includegraphics{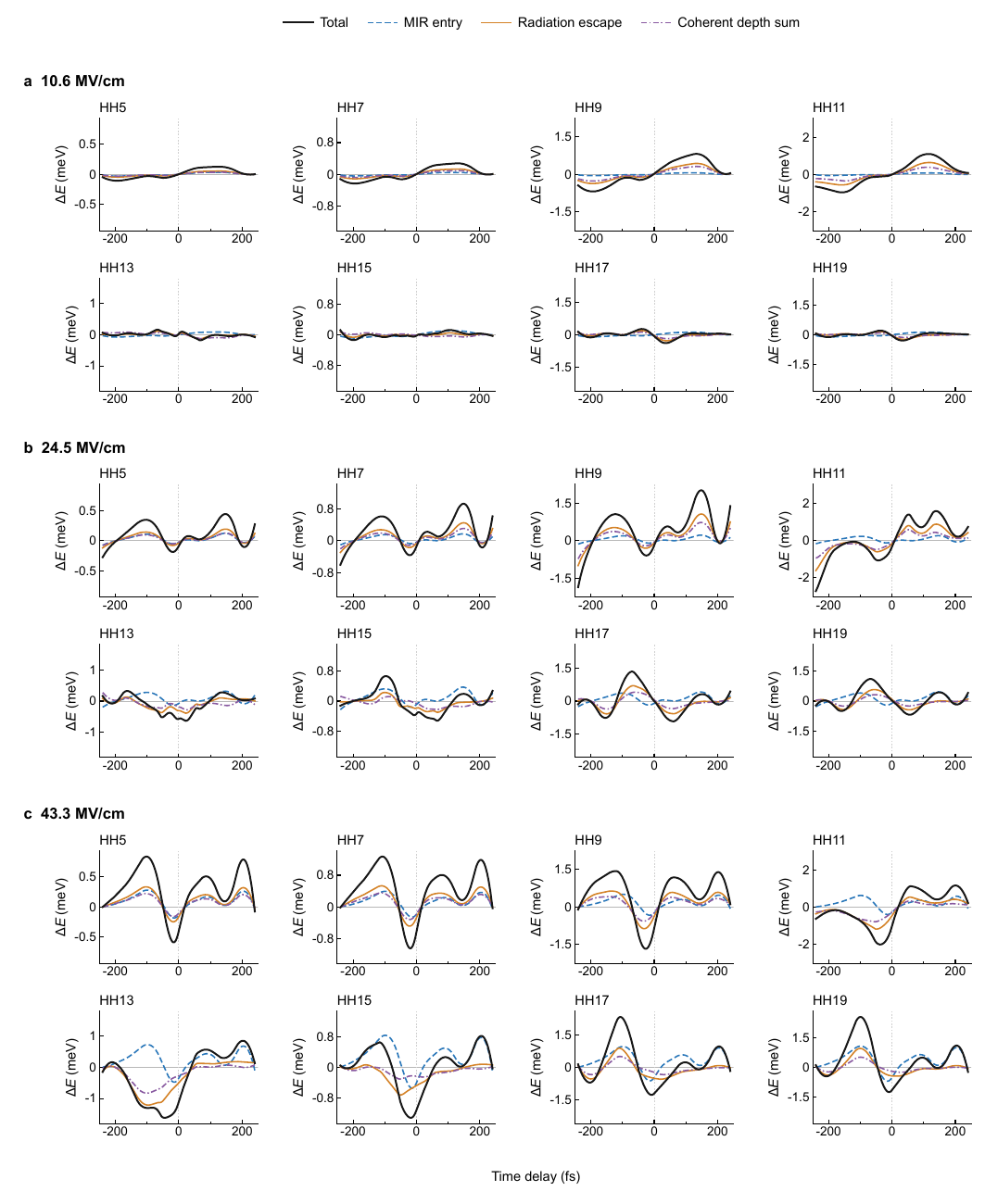}
\caption{\label{sfig:harmonic propagation shifts} Propagation-induced harmonic energy shifts. (a-c) Calculated instantaneous energy shifts for HH5–HH19 at MIR fields of 10.6, 24.5, and 43.3~MV/cm, respectively. Black curves show the total shift; blue dashed, orange solid, and purple dash-dotted curves show the contributions from MIR entry, harmonic radiation and escape, and coherent addition over depth, respectively.}
\end{figure*}

As shown in Fig.~\ref{sfig:harmonic propagation shifts}, for HH5–HH19 at all three representative driving fields, the calculated total instantaneous shifts remain on the few-meV scale, with $|\Delta E_q|<2.8$~meV over the analyzed interval. These values are much smaller than the observed harmonic shifts, which reach approximately 270~meV for HH19 (Fig.~\ref{fig:m1}d). 

We also note that the calculated curves represent instantaneous phase contributions. Since our major discovery in this work is that the harmonic emission profile in the time domain will be asymmetric with respect to pulse center, we do not impose a Gaussian temporal harmonic intensity envelope. Nevertheless, since $|\Delta E_q|$ is only a few meV throughout our driving conditions, we expect that any reasonable weighting over them will still lead to the integrated response on a similar energy scale, far below our experimentally resolved shifts. Thus, we conclude that within the present optical-transfer model, propagation cannot account for the magnitude of the measured blueshift.

\subsection{Reflectivity spectra at various MIR field strengths}
\begin{figure}
\includegraphics[width=.55\linewidth]{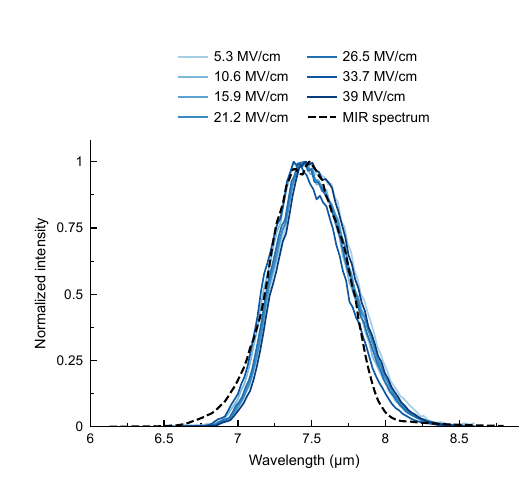}
\caption{\label{sfig:mir reflectivity} Field dependence of the reflected MIR spectrum. Peak-normalized MIR spectra reflected from Sr$_2$CuO$_3$ at the indicated driving fields (blue curves), compared with the reference MIR spectrum (black dashed curve).}
\end{figure}

To provide a complementary experimental check of the contribution from self-phase modulation, we characterized the reflected MIR spectra at driving fields from 5.3 to 39~MV/cm (Fig.~\ref{sfig:mir reflectivity}) alongside the HHG measurements. After normalization to their respective maxima, the spectra substantially overlap and show no pronounced systematic shift with increasing field. Together with the optical-constant reconstruction and phase calculation, this observation supports the conclusion that propagation-induced phase modulation within the present model is insufficient to explain the observed harmonic blueshift.

\clearpage

\section{Floquet analysis of a three-level model}
\label{sec:three-level model}
Following the recent observation of redshifted HHG from Sr$_2$CuO$_3$, which was ascribed to the Mott-gap reduction associated with metallization \cite{ikeda2025high}, it is crucial to determine whether the giant blueshift observed in our experiments could analogously arise from a transient gap widening effect due to dynamical energy level repulsion \cite{kim2014ultrafast,sie2015valley,sie2017large}. To address this, we employed Floquet theory to compute the dressed quasi-energies of the relevant low-lying states nonperturbatively. Following \cite{kishida2000gigantic,ono2004linear}, the low-energy optical physics near the Mott gap is well captured by a three-level system: the ground state $|\varphi_{\mathrm{g}}\rangle$ at $\varepsilon_{\mathrm{g}} = 0$ and the nearly degenerate odd- and even-parity Hubbard excitons $|\varphi_{\mathrm{o}}\rangle$ and $|\varphi_{\mathrm{e}}\rangle$ at $\varepsilon_{\mathrm{odd}} = 1.79$~eV and $\varepsilon_{\mathrm{even}} = 1.81$~eV. Inversion symmetry along the chain dictates the dipole selection rules, yielding a modest one-photon-allowed transition matrix element $\mu_{\mathrm{go}} = \langle\varphi_{\mathrm{g}}|\hat{x}|\varphi_{\mathrm{o}}\rangle = 0.737$~\AA{}, an anomalously large intra-exciton transition $\mu_{\mathrm{oe}} = \langle\varphi_{\mathrm{o}}|\hat{x}|\varphi_{\mathrm{e}}\rangle = 7.37$~\AA{}, and a forbidden transition $\langle\varphi_{\mathrm{g}}|\hat{x}|\varphi_{\mathrm{e}}\rangle = 0$ by parity. We note that this structure is similar to the excitonic states in another one-dimensional Mott insulator \cite{yamakawa2023dynamical}. In the dipole gauge, the driven Hamiltonian is
\begin{equation}
    \mathcal{H}(t) = \mathcal{H}_0 - e\hat{x}\,E_0\cos(\Omega t),
    \label{eq:floquet_ham}
\end{equation}
where $\hbar\Omega$ is the driving photon energy.

Assuming a periodic drive, the dressed levels are the quasi-energies of the effective Floquet Hamiltonian. Writing $|\Psi(t)\rangle = e^{-i\epsilon t/\hbar}\sum_m e^{-im\Omega t}|\varphi, m\rangle$ maps Eq.~\eqref{eq:floquet_ham} onto a static block-tridiagonal eigenvalue problem in the extended space,
\begin{equation}
    \left[\mathcal{H}_F\right]_{mm} = \mathcal{H}_0 - m\hbar\Omega\,\mathds{1},
    \qquad
    \left[\mathcal{H}_F\right]_{m,m\pm 1} = -\tfrac{1}{2}e\hat{x}E_0,
    \label{eq:sambe}
\end{equation}
which we diagonalize exactly with $|m| \le 120$ photon sectors. Because the MIR drive frequency is almost an order of magnitude below the gap ($\hbar\Omega \ll \varepsilon_{\mathrm{odd}}$), the counter-rotating (Bloch--Siegert) terms are comparable in magnitude to the co-rotating terms and must be retained. Consequently, we made no rotating-wave approximation. The model conserves the generalized parity $\hat{G} = \hat{\Pi}\,(-1)^{m}$ (spatial parity times photon-number parity), so the Floquet spectrum separates into two symmetry sectors that cross rather than repel. Each level energy is extracted as the spectral centroid of its own symmetry sector within one Floquet zone of the bare energy,
\begin{equation}
    \varepsilon_a(E_0) = \frac{\sum_{\alpha}{}' w_\alpha E_\alpha}
                              {\sum_{\alpha}{}' w_\alpha},
    \qquad
    w_\alpha = \sum_m \left|\langle a, m|\alpha\rangle\right|^2,
    \label{eq:centroid}
\end{equation}
where the primed sum runs over Floquet eigenstates $|\alpha\rangle$ with $|E_\alpha - \varepsilon_a| \le \hbar\Omega/2$ and generalized parity equal to that of the bare state $|a\rangle$.

Throughout this section, we evaluate the quasi-energies at the experimental internal field strengths (see discussions in Sec.~\ref{sec:trR}), and the highest employed amplitude is approximately 24~MV/cm. The results (Fig.~\ref{sfig:floquet shifts}) are governed by three effects with distinct field scales. First, the giant $\mu_{\mathrm{oe}}$ locks the exciton doublet into symmetric and antisymmetric dressed combinations already at fields of a few hundred kV/cm. In a frame that absorbs the strong intra-doublet drive, the splitting is renormalized by cycle-averaging to
\begin{equation}
    \Delta_{\mathrm{eff}}(E_0)
    = \Delta\, J_0\!\left(\frac{2\mu_{\mathrm{oe}} E_0}{\hbar\Omega}\right),
    \qquad \Delta = \varepsilon_{\mathrm{even}} - \varepsilon_{\mathrm{odd}},
    \label{eq:bessel}
\end{equation}
so the two exciton branches oscillate about the doublet midpoint and cross whenever the Bessel function $J_0$ changes sign (e.g., first crossing at $E_0 \simeq 1.2\,\hbar\Omega/\mu_{\mathrm{oe}} \approx 2.7$~MV/cm for $\hbar\Omega = 164$~meV, then every $\pi\hbar\Omega/2\mu_{\mathrm{oe}} \approx 3.5$~MV/cm). Second, apart from this bounded undulation, the excitons acquire essentially no net shift: for a quasi-degenerate pair the second-order optical-Stark and Bloch--Siegert shifts generated by $\mu_{\mathrm{oe}}$ cancel, so even a dressing parameter as large as $\mu_{\mathrm{oe}}E_0/\hbar\Omega \approx 11$ at 24~MV/cm leaves the pair pinned near 1.8~eV. Third, the ground state is pushed down by the weak $\mu_{\mathrm{go}}$ coupling with the usual quadratic AC-Stark form,
\begin{equation}
    \delta\varepsilon_{\mathrm{g}}
    \simeq -\left(\frac{\mu_{\mathrm{go}} E_0}{2}\right)^{\!2}
    \left(\frac{1}{\varepsilon_{\mathrm{odd}} - \hbar\Omega}
        + \frac{1}{\varepsilon_{\mathrm{odd}} + \hbar\Omega}\right),
    \label{eq:acstark}
\end{equation}
reaching $-10$ to $-20$~meV at 20~MV/cm. At still higher fields the curves turn around sharply as the dressed $\mathrm{g}\!\to\!\mathrm{o}$ gap sweeps through an $N$-photon resonance $N\hbar\Omega$ (with $N = 11, 10, 9, 8$ for $\hbar\Omega = 164, 180, 200, 220$~meV), and as the resonance is crossed the sign of the level repulsion between the ground ladder and the exciton ladder flips, converting the ground-state downshift into a steep upshift near 22-24~MV/cm.

\begin{supplementaryfigure}
\includegraphics{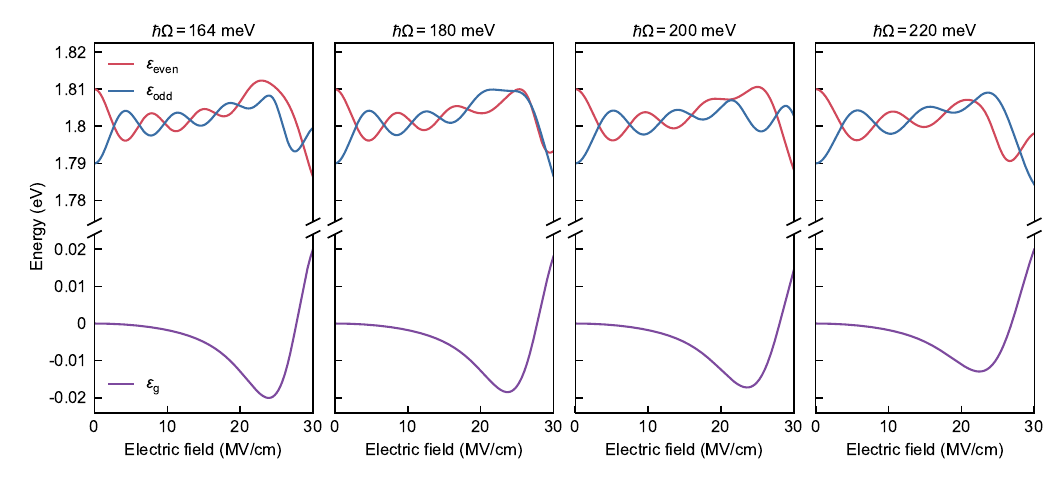}
\caption{\label{sfig:floquet shifts} Floquet quasi-energies of the driven three-level system. Dressed energies of the even exciton ($\varepsilon_{\mathrm{even}}$, red), odd exciton ($\varepsilon_{\mathrm{odd}}$, blue), and the ground state ($\varepsilon_{\mathrm{g}}$, purple) as a function of the peak field strength $E_0$ for four driving photon energies $\hbar\Omega = 164$, 180, 200, and 220~meV.}
\end{supplementaryfigure}

The quasi-energies of Fig.~\ref{sfig:floquet shifts} characterize a strictly periodic drive. To describe the dynamics in pulse-laser-driven systems, we adopt the instantaneous-Floquet-state picture \cite{breuer1989quantum,drese1999floquet,holthaus2016floquet}: when the envelope contains several optical cycles, the system adiabatically follows the Floquet states of the instantaneous field amplitude, so the levels evolve in real time as $\varepsilon_a(t) = \varepsilon_a(E_0(t))$. Numerical simulations of state populations through the time-dependent Schrödinger equation show that this adiabatic reading remains quantitatively accurate down to pulses of roughly two cycles \cite{ikeda2022floquet}. Thus, our pulse parameters (100~fs intensity FWHM, i.e.\ 4--5.3 cycles across $\hbar\Omega$ = 164--220~meV) lie safely within this regime. Since we do not focus on the carrier-envelope-phase dependence of the dynamics, below we analyze the time evolution of the energy levels of the three-state model with a cosine-phase carrier wave.

Figure~\ref{sfig:floquet dynamics} shows the resulting level dynamics for the experimental peak field of 24~MV/cm. As the envelope rises, the exciton doublet sweeps through the sequence of $J_0$ crossings of Eq.~\eqref{eq:bessel} before collapsing onto the doublet midpoint, where the pair remains pinned during the central $\sim$50~fs of the pulse. The ground state follows a smooth AC-Stark dip, reaching $-20$~meV at $\hbar\Omega = 164$~meV; at 220~meV the dip develops a shallow double minimum as the dressed gap grazes the 8-photon resonance near the peak. Because the harmonic emission energy is related to the level shift during the sub-cycle carrier propagation, we sample the transition energies $\delta\varepsilon_{\mathrm{odd}} = \varepsilon_{\mathrm{odd}} - \varepsilon_{\mathrm{g}}$ and $\delta\varepsilon_{\mathrm{even}} = \varepsilon_{\mathrm{even}} - \varepsilon_{\mathrm{g}}$ at the crest times $t_k = k\,T/2$, where the instantaneous field magnitude is locally maximal. The gaps open toward the blue as the pulse builds, by at most $+38$~meV ($\mathrm{g}\!\to\!\mathrm{o}$) and $+22$~meV ($\mathrm{g}\!\to\!\mathrm{e}$) at $\hbar\Omega = 164$~meV (the third row in Fig.~\ref{sfig:floquet dynamics}). The bottom row of Fig.~\ref{sfig:floquet dynamics} resolves the crest-to-crest variation $\Delta\delta\varepsilon(t_k) = \delta\varepsilon(t_{k+1}) - \delta\varepsilon(t_k)$: the gap drifts by at most $\sim$11~meV per half cycle, with a slow component that is positive on the rising edge and negative on the falling edge and a fast sign-alternating component on the pulse flanks inherited from the $J_0$ crossings. This sub-cycle modulation is more than an order of magnitude too small to account for the giant, hundreds-of-meV blueshifts observed experimentally. Therefore, we rule out the sub-cycle variation of optical transitions between the ground and the excitonic states as the origin of the non-integer HHG.

\begin{supplementaryfigure}
\includegraphics[width=\textwidth]{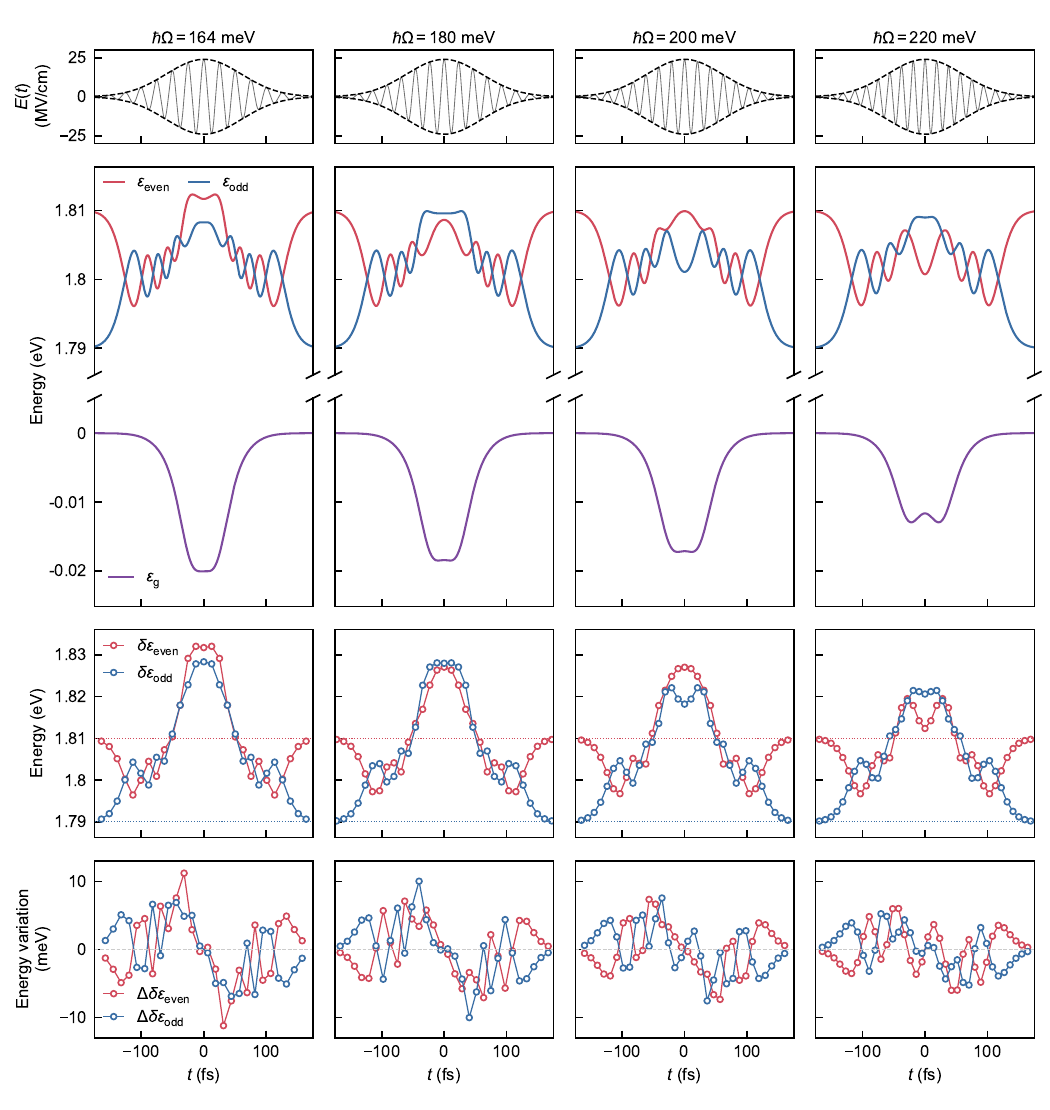}
\caption{\label{sfig:floquet dynamics} Real-time Floquet level dynamics during the pulsed excitation. Top row: driving field $E(t) = E_0(t)\cos(\Omega t)$ (solid) and envelope $\pm E_0(t)$ (dashed) for a Gaussian pulse with 100~fs intensity full width at half maximum (FWHM) and 24~MV/cm peak field, for the four driving photon energies. Second row: instantaneous Floquet levels $\varepsilon_a(E_0(t))$ of the even exciton ($\varepsilon_{\mathrm{even}}$, red), odd exciton ($\varepsilon_{\mathrm{odd}}$, blue), and the ground state ($\varepsilon_{\mathrm{g}}$, purple). Third row: dynamic transition energies $\delta\varepsilon = \varepsilon_{\mathrm{even/odd}} - \varepsilon_{\mathrm{g}}$ sampled at the field crests $t_k = k\,T/2$ (twice per optical cycle); dotted lines mark the bare transition energies at 1.79 and 1.81~eV. Bottom row: change of the transition energies between consecutive crests, $\Delta\delta\varepsilon = \delta\varepsilon(t_{k+1}) - \delta\varepsilon(t_k)$, giving the transition energy drift accumulated per half cycle.}
\end{supplementaryfigure}

\clearpage

\section{Optical Bloch equations simulations of HHG from the excitonic states}
\label{sec:exciton hhg}
Because the optical properties of Sr$_2$CuO$_3$ near the Mott gap edge are dominated by the excitonic transitions (Sec.~\ref{sec:carrier density}), we further utilize the same three-level model developed in Sec.~\ref{sec:three-level model} to exclude the origin of the observed non-integer HHG from the coupling between the ground and excitonic states. To this end, we simulate the charge dynamics using the optical Bloch equations,
\begin{equation}
    \dot{\rho} = -\frac{i}{\hbar}\left[\mathcal{H}(t), \rho\right] + \mathcal{D}[\rho],
    \label{eq:obe}
\end{equation}
where $\rho$ is the density matrix of the three-level system, $\mathcal{H}(t)$ is the Hamiltonian of Eq.~\eqref{eq:floquet_ham} with $E(t) = E_0(t)\cos(\Omega t)$ (Gaussian envelope, 100~fs intensity FWHM), and the dissipator $\mathcal{D}$ damps all coherence at the dephasing rate $1/T_2$ and relaxes both exciton populations to the ground state at $1/T_1$. We use $T_1 = 2$~ps from the measured exciton lifetime \cite{ogasawara2000ultrafast} and $T_2 = 21.8$~fs (approximately one optical cycle of the drive). The emitted spectrum is computed directly from the time-dependent polarization $P(t) = \mathrm{Tr}[\rho(t)\,e\hat{x}]$ as $S(\omega) \propto \omega^4 |P(\omega)|^2$. We verified that the features discussed below are robust against the choice of $T_2$: fully coherent propagation ($T_2 \to \infty$) adds spectral interference fringes but does not move the emission bands, while a shorter dephasing time ($T_2 = T/4$) only broadens them.

Figure~\ref{sfig:obe hhg} shows the calculated emission spectra as a function of the peak driving field from the perturbative regime up to the experimental field strength for the four photon energies. Across all fields and frequencies, the emission is organized into odd harmonics of the drive (dashed lines), as enforced by the dynamical inversion symmetry of the driven parity-conserving model. While the harmonics broaden and develop complex substructure at high field strengths, their central frequencies remain rigidly anchored to the integer grid across the entire 0 to 4~eV spectral range. The only systematic off-integer spectral weight appears in the immediate vicinity of the exciton resonance at $\sim 1.8$~eV. More importantly, there is no cumulative drift of the harmonic comb with increasing field. The shifts neither grow across the broad spectral range nor accumulate progressively with harmonic order. Together, these results demonstrate that coupling among discrete excitonic states cannot produce the giant, order-dependent non-integer blueshifts observed in our experiments.

\begin{supplementaryfigure}
\includegraphics{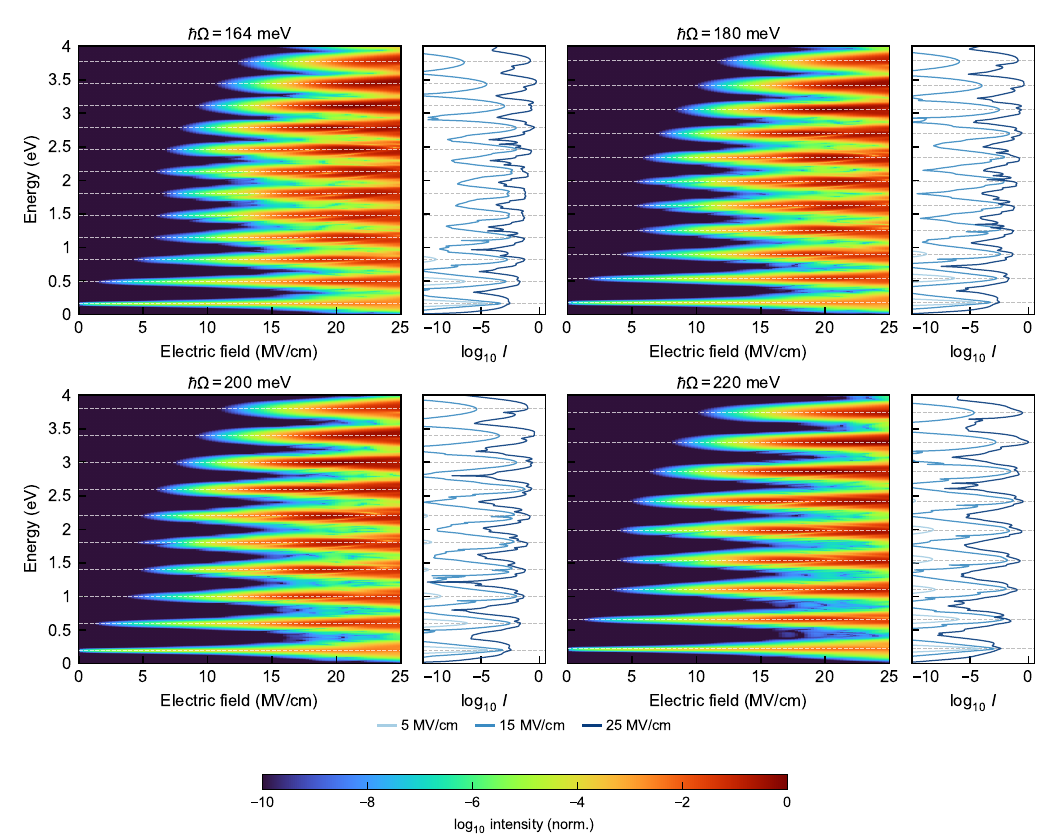}
\caption{\label{sfig:obe hhg} HHG from the driven three-level model. Emission spectra computed from the optical Bloch equations as a function of the peak electric field for driving photon energies $\hbar\Omega = 164$, 180, 200, and 220~meV. Color maps show the normalized emission intensity on a logarithmic scale as a function of peak driving field and emission energy; dashed lines mark the odd-integer multiples of $\hbar\Omega$. The narrow side panels display spectral linecuts at peak fields of 5, 15, and 25~MV/cm (light to dark blue). With increasing field, emission at successively higher harmonics switches on, and the peaks broaden and develop substructure while remaining organized around the odd-harmonic energies.}
\end{supplementaryfigure}

\clearpage

\section{Nonadiabatic phase in HHG}
\label{sec:nonadiabatic phase}
In this section, we derive the origin of the nonadiabatic phase in HHG by recalling the atomic dipole in the Lewenstein form and generalizing this framework to a crystalline lattice through the interband and intraband currents in a closed form.

Throughout this section, we use atomic units ($\hbar=e=m_e=1$); the carrier angular frequency is $\Omega$, the optical period $T=2\pi/\Omega$, and the nominal $q$th harmonic sits at $q\Omega$. The laser is defined as a linearly polarized Gaussian pulse with electric field $\mathbf{E}(t)=-\,d\mathbf{A}/dt$ and vector potential $\mathbf{A}(t)$. Crystal momenta are referred to the Houston frame $\mathbf{k}\!\to\!\mathbf{k}+\mathbf{A}(t)$.

\subsection{Atomic HHG}
\label{sec:SM-lewenstein}
In a gas, the strong-field dipole is built from three steps: tunnel ionization of the bound electron, field-driven propagation of the free electron in the continuum, and radiative recombination of the electron with the parent ion. In the single-active-electron and strong-field limits, the time-dependent dipole $\mathbf{x}(t)$ is given by the Lewenstein integral \cite{lewenstein1995phase}
\begin{equation}
\mathbf{x}(t)=i\!\int_{-\infty}^{t}\!\! dt'\!\!\int\! d^3\mathbf{p}\;
   \mathbf{d^{*}}\!\big[\mathbf{p}+\mathbf{A}(t)\big]\,
   \mathbf{E}(t')\,\mathbf{d}\big[\mathbf{p}+\mathbf{A}(t')\big]\,
   e^{-iS(\mathbf{p},t',t)}+\text{c.c.},
\label{eq:SM-lewdipole}
\end{equation}
where $\mathbf{p}$ is the canonical momentum conserved during the excursion, $\mathbf{d}(v)=\langle v|\mathbf{x}|0\rangle$ is the bound--continuum transition dipole moment, and the quasiclassical action is
\begin{equation}
S(\mathbf{p},t',t)=\int_{t'}^{t}\!\Big[\tfrac12\big(\mathbf{p}+\mathbf{A}(\tau)\big)^2   +I_p\Big]\,d\tau .
\label{eq:SM-ataction}
\end{equation}
The three steps appear as the three factors in Eq.~\eqref{eq:SM-lewdipole}: $\mathbf{E}(t')\,\mathbf{d}[\mathbf{p}+\mathbf{A}(t')]$ is the ionization amplitude at the birth time $t'$, $e^{-iS}$ is the propagation phase, and $\mathbf{d^{*}}[\mathbf{p}+\mathbf{A}(t)]$ is the recombination amplitude at time $t$. The harmonic spectrum is the Fourier transform of the dipole, $x(\omega)=\int dt\,e^{i\omega t}x(t)$, whose integrand carries a rapidly varying phase $\Phi(\mathbf{p},t',t)=\omega t-S(\mathbf{p},t',t)$. We can express the $q$-th order contribution to the emitted field as $x_q(t)=a_q(t)\,e^{\,i\Phi_q(t)}$ with $\Phi_q(t)=q\Omega t+\theta_q(t)$, where the intrinsic harmonic phase is the stationary value of the action stripped of the carrier reference,
\begin{equation}
\theta_q(t)=S\big(\mathbf{p},t'_s,t\big)-q\Omega t .
\label{eq:SM-thetaq}
\end{equation}
The instantaneous emission frequency is therefore $\omega(t)=d\Phi_q/dt=q\Omega+\dot\theta_q$. Crucially, the deviation from the integer comb line is \begin{equation}
\Delta\omega_q(t)=\dot\theta_q
=\Big(\frac{d\theta_q}{dI}\Big)\dot I
=2I\Big(\frac{d\theta_q}{dI}\Big)\frac{\dot E}{E}.
\label{eq:SM-chirp}
\end{equation}
This equation reveals that the nonadiabatic harmonic frequency shift is expressed with two terms: the intensity-dependent harmonic phase and the relative electric-field variation rate. Consequently, it vanishes at the pulse peak, becomes positive (a blueshift) on the rising edge ($\dot E>0$) and negative (a redshift) on the falling edge ($\dot E<0$). Furthermore, the latter term also indicates that the nonadiabatic phase generally accumulates more rapidly with longer driving wavelengths and shorter pulse durations, which was demonstrated in Ar and Ne atoms \cite{shin2001nonadiabatic}.

\subsection{Solid-state HHG}
\label{sec:nonadiabatic solid-state hhg}
In a crystalline solid, the continuum is replaced by conduction bands and the bound state by valence bands. Nevertheless, the quasiclassical action can be analogously defined by integrating the momentum-dependent bandgap $\varepsilon_g$ along the Houston-frame trajectory \cite{vampa2014theoretical}
\begin{equation}
    S(\mathbf{k},t,t')=\int_{t'}^{t}\varepsilon_g\big(\mathbf{k}+\mathbf{A}(\tau)-\mathbf{A}(t)\big)d\tau.
\end{equation}
Then, the polarization is given by
\begin{equation}
    \mathbf{P}(\mathbf{k},t)=d\big(\mathbf{k}+\mathbf{A}(t)\big)\,\rho_{cv}(\mathbf{k},t)\,e^{iS(\mathbf{k},t)}+\text{c.c.},
    \label{eq:SM-sHHG-pol}
\end{equation}
where $\mathbf{d}$ is the transition dipole moment and $\rho_{cv}$ is the interband coherence. Using the Keldysh approximation $n_v-n_c\approx1$ \cite{keldysh1964ionization}, the two-level density matrix equations can be decoupled and formally integrated, giving rise to analytical expressions for the interband and intraband currents
\begin{align}
\mathbf{j}_{\rm inter}(\omega)=&\omega\!\int d^3\mathbf{k}\,\mathbf{d}(\mathbf{k})\!\int_{-\infty}^{\infty}\!\!dt\,e^{-i\omega t}\!\int_{-\infty}^{t}\!\!dt'\,\mathbf{E}(t')\,\mathbf{d}^{*}\big(\mathbf{k}+\mathbf{A}(t')-\mathbf{A}(t)\big)\,e^{-iS(\mathbf{k},t',t)-(t-t')/T_2}+\text{c.c.},
\label{eq:SM-4a}\\
\mathbf{j}_{\rm intra}(\omega)=&\sum_{m}s_m\int d^3\mathbf{k}\,\mathbf{v_m}(\mathbf{k})\!\int_{-\infty}^{\infty}\!\!dt\,e^{-i\omega t}\!\int_{-\infty}^{t}\!\!dt'\,\mathbf{E}(t')\,\mathbf{d}\big(\mathbf{k}+\mathbf{A}(t')-\mathbf{A}(t)\big)\!\nonumber\\
&\times\int_{-\infty}^{t'}\!\!dt''\,\mathbf{E}(t'')\,\mathbf{d^{*}}\big(\mathbf{k}+\mathbf{A}(t'')-\mathbf{A}(t)\big)\,e^{-iS(\mathbf{k},t'',t')-(t'-t'')/T_2}+\text{c.c.} \nonumber\\&\text{($s_m$ = $-1$, 1 for $m$ = $v$, $c$)}
\label{eq:SM-4b}
\end{align}
Because the mathematical structure of Eqs.~\eqref{eq:SM-4a} and \eqref{eq:SM-4b} highly mirrors the Lewenstein dipole (Eq.~\eqref{eq:SM-lewdipole}), the nonadiabatic phase mechanism derived for atomic gases will also directly apply to solids.

\begin{figure}[b]
\includegraphics{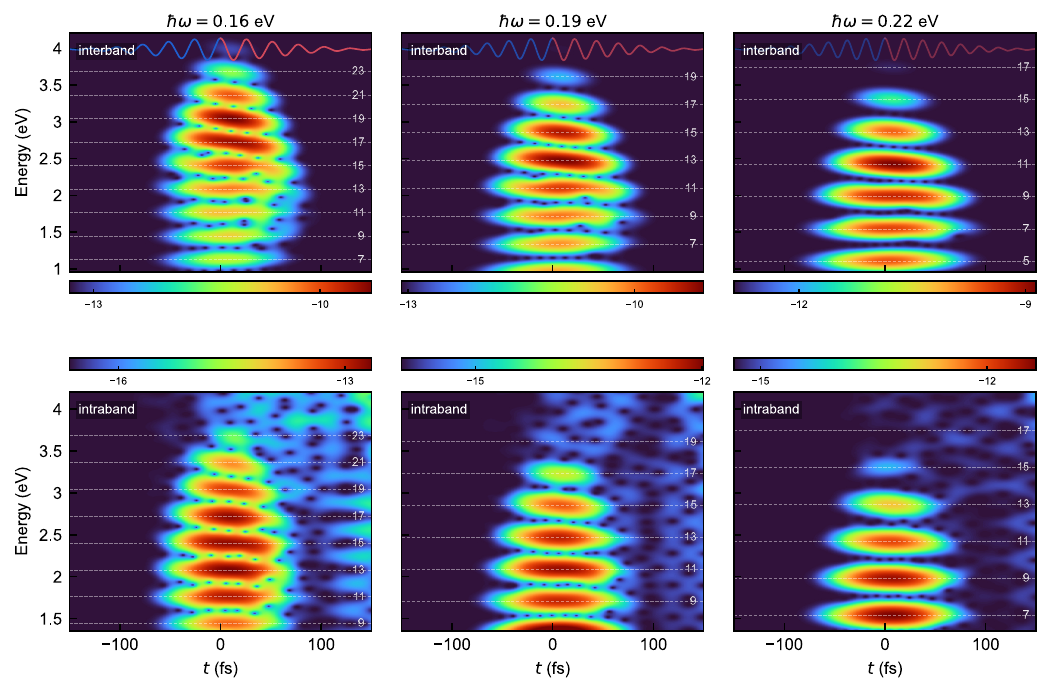}
\caption{\label{sfig:nonadiabatic phase} Time-frequency structure of the interband and intraband emission for three drive photon energies. Gabor spectrograms of the interband (top row) and intraband (bottom row) currents evaluated for the holon dispersion in the one-dimensional Hubbard model. The spectrograms are computed by Fourier transforming the current within a sliding Gaussian window of 24~fs FWHM. Columns correspond to drive photon energies $\hbar\omega=0.16$, $0.19$ and $0.22$~eV. Color encodes $\log_{10}$ of the spectral intensity. White dashed lines mark the odd-integer harmonic energies $n\hbar\omega$ and are labeled by the harmonic order $n$. All runs use the same peak field.}
\end{figure}

To visually demonstrate the effect of this nonadiabatic phase, we numerically solved the intraband and interband currents utilizing the exact holon dispersion of the one-dimensional Hubbard model (Eqs.~\eqref{eq:holon exp1} and \eqref{eq:holon exp2}, see details in Sec.~\ref{sec:sbe}). We use a constant dephasing time $T_2=T/6$ in order to visualize the structure of each harmonic. Figure~\ref{sfig:nonadiabatic phase} resolves the harmonic emission in both time and energy via Gabor spectrograms, separated into its interband and intraband contributions, for three photon energies spanning the experimentally relevant range with the same peak electric field strength. The spectrograms clearly reveal a dynamic frequency chirp: the HH emission evolves from blueshifted energies on the rising edge of the pulse to redshifted energies on the falling edge. In addition, consistent with the $\dot E/E$ dependence, the amplitude of the frequency sweep decreases with increasing photon energy, mirroring the experimental trend (Fig.~\ref{fig:m3}). We emphasize that while this single-particle framework successfully captures the intrinsic nonadiabatic chirp, it produces symmetric emission on both sides of the pulse, resulting in a net cancellation. As detailed in the main text and Secs.~\ref{sec:ed} and ~\ref{sec:sbe}, it is the many-body electronic state reconstruction that breaks this temporal symmetry by abruptly quenching the falling-edge emission, thereby exposing the rising-edge nonadiabatic phase as a macroscopic, non-integer blueshift.

The increase of the blueshift with harmonic order (Figs.~\ref{fig:m1}d and~\ref{fig:m3}) can be understood qualitatively from the excursion times of doublon–holon pairs. Along the short-trajectory branch, higher-order harmonics generally originate from longer intervals between pair creation and recombination (Fig.~\ref{sfig:order dependence}). For comparable envelope slopes, these trajectories sample a larger change in the driving-field amplitude during propagation, which can enhance the intensity sensitivity of the accumulated harmonic phase and hence the nonadiabatic frequency shift in Eq.~\eqref{eq:SM-chirp}. This trajectory picture is supported by recent two-color HHG measurements in another one-dimensional cuprate Mott insulator, where the observed positive emission chirp was interpreted as evidence for predominantly short doublon–holon recombination trajectories \cite{hatch2026sub}.

\begin{supplementaryfigure}
\includegraphics{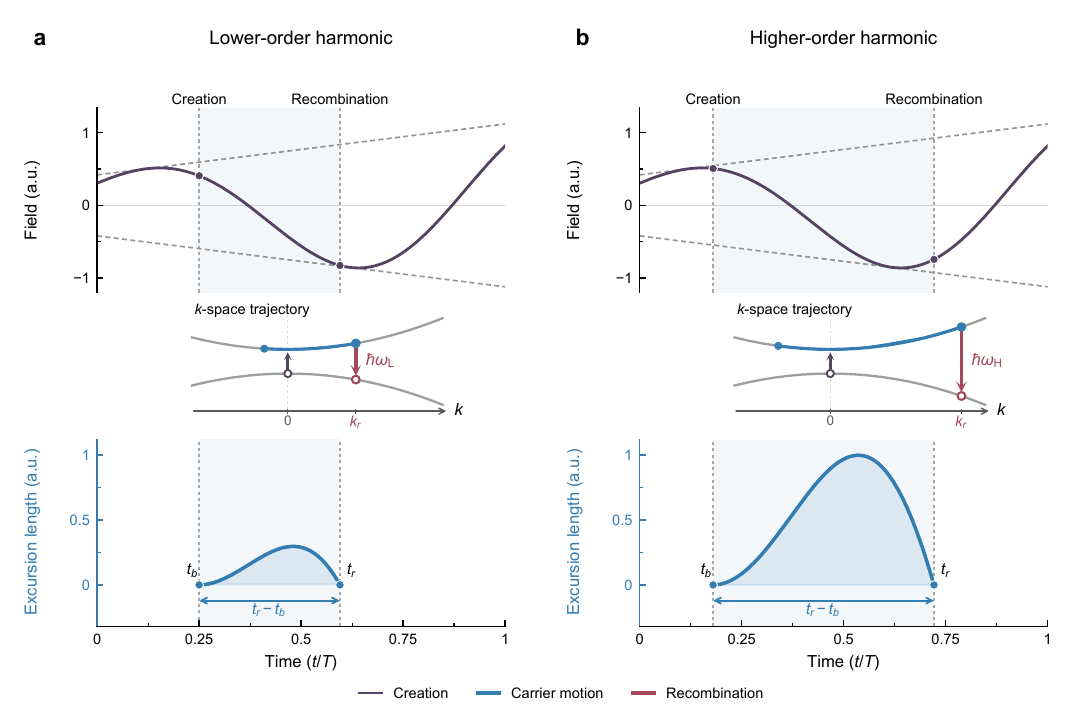}
\caption{\label{sfig:order dependence} Schematic origin of order-dependent spectral shifts in interband HHG dominated by short trajectories. Both panels show trajectories belonging to the short branch, producing a lower-order harmonic (a) and a higher-order harmonic (b). Upper traces show the same driving-field cycle on a rising pulse envelope (gray dashed lines), with vertical guides marking pair creation $t_b$ and recombination $t_r$. Middle sketches show identical two-band dispersions: purple arrows indicate creation, blue arrows indicate motion in momentum space, and red arrows denote recombination, releasing photon energies $\hbar\omega_{\mathrm{L}}$ and $\hbar\omega_{\mathrm{H}}$. Lower traces show the relative pair separation; double-headed arrows indicate the excursion duration $t_r-t_b$. Time is normalized to the optical period (T); the remaining scales are arbitrary.}
\end{supplementaryfigure}

\clearpage

\section{Time-dependent exact diagonalization simulations}
\label{sec:ed}
To rigorously capture the real-time evolution of the strongly correlated many-body state and the associated microscopic currents, we performed time-dependent exact diagonalization (ED) simulations on a single-band one-dimensional extended Fermi-Hubbard model. Based on the quantitative experiment-theory comparisons in \cite{chen2021anomalously,wang2021phonon}, this minimal model can faithfully describe essential physical properties of the quasi-one-dimensional cuprate Sr$_2$CuO$_3$, with its Hamiltonian given by:
\begin{align}
    \mathcal{H}(t)=-t_h\sum_{j\sigma}[c_{j\sigma}^\dagger c_{j+1\sigma}+c_{j+1\sigma}^\dagger c_{j\sigma}] + U\sum_{j}n_{j\uparrow}n_{j\downarrow} + V\sum_{i\sigma\sigma'}n_{i\sigma}n_{i+1,\sigma'}.
\end{align}
Here, $c_{j\sigma}^\dagger$ ($c_{j\sigma}$) creates (annihilates) an electron at site $j$ with spin $\sigma$. The parameter $t_h$ is the nearest-neighbor hopping integral and $U$ is the on-site Coulomb repulsion. $U=8t_h$ and $V=-t_h$ are set to simulate the strong correlation and near-neighbor attraction in this system. In order to confirm that the nonadiabatic gating mechanism originates from the strong correlations in the model, we also performed time-dependent ED simulations in pure Hubbard models without additional terms and reproduced the same phenomenology, as detailed below in Secs.~\ref{sec:1d hubbard} and \ref{sec:2d hubbard}. The pulsed laser field is simulated with a time-dependent electric field (vector potential)
\begin{equation}
    A(t) = A_0 f(t)\cos{\Omega(t-t_0)},
\end{equation}
where $f(t)=e^{-(t-t_0)^2/2\sigma_t^2}$ is the Gaussian envelope function of the pulse. To match the experimental excitation conditions, we employ a Gaussian pulse envelope with an intensity FWHM of 100~fs.

We use the parallel Arnoldi method \cite{lehoucq1998arpack} to determine the ground-state wavefunction \cite{jia2018paradeisos} and the Krylov subspace technique \cite{balzer2011krylov,park1986unitary,hochbruck1997krylov,moler2003nineteen} to evaluate the evolution of a state $|\psi(t+\delta t)\rangle=e^{-i\mathcal{H}(t)\delta t}|\psi(t)\rangle$. The pump laser field enters the Hamiltonian via a Peierls' substitution. We confirm the convergence of real-time propagation by comparing the current and HHG spectra with increasing Krylov subspace dimensions and decreasing time steps.

To compute the harmonic emission, we first use the electric current operator defined as
\begin{equation}
    \hat{J}(t) = -it_h\sum_{j\sigma}[e^{-iA(t)}c_{j\sigma}^\dagger c_{j+1\sigma}-e^{iA(t)}c_{j+1\sigma}^\dagger c_{j\sigma}]
\end{equation}
to calculate the time-dependent current. HHG spectrum from charge acceleration is then given by
\begin{equation}
    S(\omega) = |\int \partial_t \langle J(t)\rangle e^{-i\omega t}dt|^2.
\end{equation}
We also evaluate the ground-state fidelity \(F(t)=|\langle\psi_\mathrm{GS}|\psi(t)\rangle|^2\) to quantify the departure from the initial Mott-insulating state. Here, \(|\psi_\mathrm{GS}\rangle\) is the normalized ground state of the field-free Hamiltonian (\(A=0\)), which initializes the evolution and remains the fixed reference state. At each propagation step, we calculate the modulus squared of its overlap with \(|\psi(t)\rangle\) under unitary evolution, without phenomenological damping or dephasing.

\subsection{System size dependence of the microscopic model}
Because computational constraints limit ED simulations to finite clusters, we first investigated the charge dynamics in 12-, 14-, and 16-site periodic chains to ensure our results are robust against finite-size effects.

As shown in Fig.~\ref{sfig:system size dependence}, all three chains are driven by the same vector potential $A(t)$ (Fig.~\ref{sfig:system size dependence}a, top), and the induced currents $J(t)$ follow nearly identical temporal profiles: as the field ramps up, they develop rapid high-frequency oscillations prior to the pulse maximum, and cease ringing at the trailing edge of the excitation (Fig.~\ref{sfig:system size dependence}a). The corresponding HH emission spectra are overlaid in Fig.~\ref{sfig:system size dependence}b. The three spectra nearly coincide over the entire range, reproducing the same harmonic plateau and high-order cutoff, which aligns with the energy of the Hubbard bands \cite{silva2018high,murakami2021high,orthodoxou2021high,huang2023quasiparticle}; only the fine structure of individual peaks varies slightly with system size, as expected from the discrete level spacing of finite clusters. The close agreement of both the real-time currents and the emission spectra demonstrates that the 12-site chain already captures the essential charge dynamics and HHG responses, validating its use throughout this work.

\begin{supplementaryfigure}
\includegraphics[width=.68\linewidth]{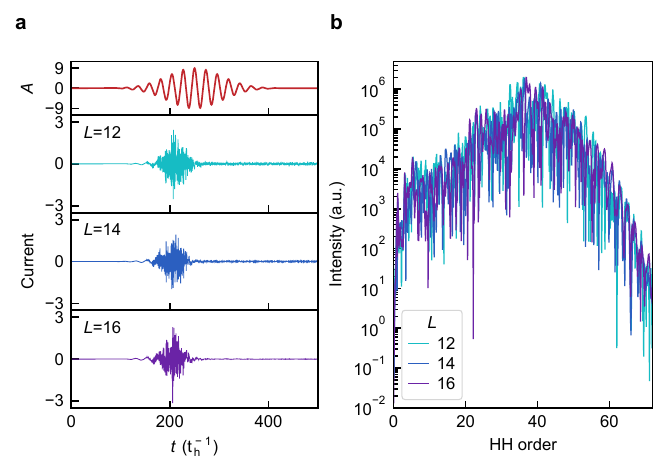}
\caption{\label{sfig:system size dependence} Finite-size convergence of the time-dependent ED HHG simulations. (a) Driving vector potential $A(t)$ (top, red) and the resulting induced current $J(t)$ for 12-, 14-, and 16-site extended Hubbard chains (shown in separate panels), as a function of time (units of $\hbar/t_h$). (b) The corresponding HH emission spectra. The peak amplitude $A_0=9$, central frequency $\Omega=0.273t_h$ and a Gaussian envelope are used here.}
\end{supplementaryfigure}

\subsection{Non-perturbative nature of HHG from the extended Hubbard model}
\begin{figure*}[htb]
\includegraphics[width=.58\linewidth]{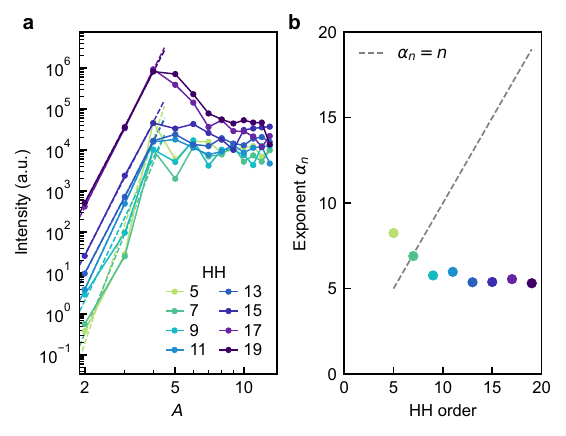}
\caption{\label{sfig:hubbard field scaling} Field-strength scaling of HHG from the Hubbard chain. (a) Integrated intensities of the odd harmonics HH5-HH19 versus $A_0$ on log-log axes. The dashed lines are power-law fits $I_n\propto A_0^{2\alpha_n}$ to the three lowest-field points ($A_0=2,3,4$). (b) Extracted power-law fitting exponents $\alpha_n$ versus harmonic order $n$ ($\alpha_5=8.3$, $\alpha_7=6.9$, $\alpha_9=5.8$, $\alpha_{11}=6.0$, $\alpha_{13}=5.4$, $\alpha_{15}=5.4$, $\alpha_{17}=5.6$, $\alpha_{19}=5.3$), compared with the perturbative expectation $\alpha_n=n$ (dashed line).}
\end{figure*}

To verify that our model operates in the same highly non-perturbative regime as the experiment, we extracted the integrated intensity $I_n$ of each odd harmonic from the driven extended Hubbard chain as a function of the peak vector potential amplitude $A_0=2$--$13$ (Fig.~\ref{sfig:hubbard field scaling}a). As shown in Fig.~\ref{sfig:hubbard field scaling}b, the simulated exponents $\alpha_n$ do not follow the expectation of perturbative power-law scaling. Instead, they remain roughly constant and lie far below $n$ for the high orders ($n>7$). $\alpha_n$ is approximately order-independent and does not obey $\alpha_n=n$. This flat, sub-linear exponent profile establishes the strongly non-perturbative nature of the generated HHs.

\subsection{Time-domain HHG spectrogram analysis}
\label{sec:subcycle analysis}
To directly visualize the truncated emission in the strong-field regime, we analyze the time-domain structure of the simulated HHG. We performed a decomposition of the current $J(t)$ by wavelet analysis to extract the temporal emission amplitude at each frequency. The time-domain HHG is computed by the continuous wavelet transform
\begin{equation}
    I(s,\tau)=\frac{1}{\sqrt{|s|}}\int_{-\infty}^{\infty} I(t)\,
    \psi^*\!\left(\frac{t-\tau}{s}\right)dt ,
\end{equation}
where $I(t)=\partial_t\langle J(t)\rangle$ is the charge acceleration used to calculate the HHG spectrum and $\psi$ is the wavelet. We used a Morlet wavelet
to perform the analysis,
\begin{equation}
    \psi(u)=\pi^{-\frac{1}{4}}\,
    e^{i\omega_0u}\,e^{-\frac{u^2}{2}} ,
\end{equation}
where $\omega_0=250$ (the dimensionless Morlet central frequency) is adopted to ensure simultaneously sufficient time and frequency resolution. The scale $s$ maps onto emission frequency through $\omega=\omega_0/s$, which we express as the harmonic order $n=\omega/\Omega$ relative to the drive frequency $\Omega=0.273\,t_h$.

\begin{figure*}[hb]
\includegraphics{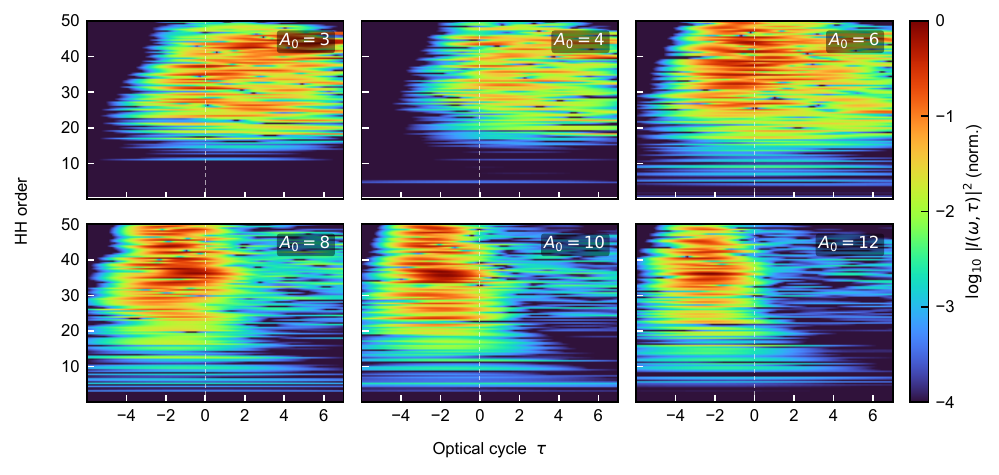}
\caption{\label{sfig:subcycle analysis} Time-domain wavelet analysis of the simulated HH emission. Each panel is individually normalized to its own maximum.}
\end{figure*}

The resulting time-frequency emission maps $|I(s,\tau)|^2$ are shown in Fig.~\ref{sfig:subcycle analysis} for peak amplitudes $A_0=3$--$12$. At the lowest fields ($A_0=3,4$), the harmonic emission is distributed around and after the pulse center, with substantial intensity persisting on the trailing edge---a consequence of the long-lived residual current that, in the absence of decoherence channels in the model, continues to radiate after the pulse \cite{murakami2021high}. However, as the field strength increases into the strong-field regime ($A_0\geq6$), this trailing-edge emission is progressively and severely suppressed and the dominant emission window shifts systematically toward earlier optical cycles on the rising edge of the pulse. By $A_0=10$, essentially all of the harmonic emission is confined to the leading edge. This temporal gating mirrors our phenomenological framework in Sec.~\ref{sec:nonadiabatic phase}: once the intense field drives the system across the intra-pulse electronic state reorganization, the rapid photo-generated doublon-holon population metallizes the chain and destroys the coherence between the many-body wavefunction and the ground state required for sustained HHG. This abruptly truncates the emission window, yielding the asymmetric, rising-edge-biased harmonics that macroscopically manifest as the non-integer blueshift.

\subsection{Ground-state fidelity and temporal gating}
\label{sec:hubbard fidelity}
\begin{figure}[hb]
\includegraphics{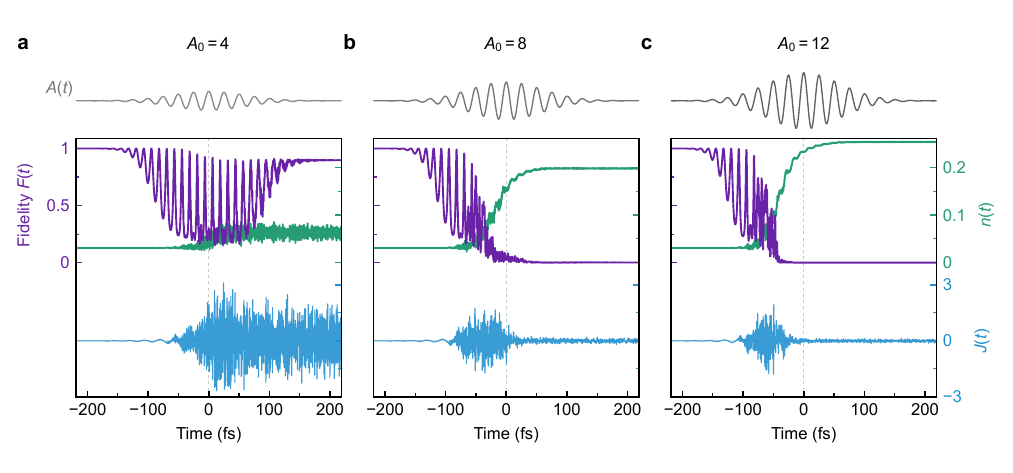}
\caption{\label{sfig:fidelity} Field-dependent many-body dynamics in the driven extended Hubbard chain. Time-dependent exact diagonalization simulation results for peak fields $A_0$ of (a), 4; (b), 8; and (c), 12. Upper traces show the driving vector potential on a common amplitude scale. Below, synchronized ground-state fidelity \(F(t)\) (purple), doublon density per site \(n(t)\) (green), and the current \(J(t)\) (blue) are displayed. Vertical dashed lines mark the pulse center.}
\end{figure}

To examine the many-body dynamics underlying the temporal gate, we compare the ground-state fidelity with the induced current. At \(A_0=3,4\), the fidelity exhibits pronounced oscillations during the pulse but substantially recovers afterwards, reaching approximately 0.99 and 0.9 at the end of the simulation (Figs.~\ref{fig:m4}a and~\ref{sfig:fidelity}a). Increasing the driving amplitude progressively suppresses this recovery: the final fidelity falls to approximately 0.12 for \(A_0=6\) (Figs.~\ref{fig:m4}b) and below \(2\times10^{-3}\) for \(A_0\geq8\) (Fig.~\ref{sfig:fidelity}b). At even stronger driving amplitudes, the ground-state weight is already strongly depleted before the pulse maximum (Figs.~\ref{fig:m4}c and~\ref{sfig:fidelity}c). This evolution accompanies the suppression of the trailing-edge current and the concentration of harmonic emission toward the rising edge identified by the spectrogram analysis.

The connection to ground–excited-state interference can be expressed by decomposing the wavefunction as \(|\psi(t)\rangle=a_{\mathrm{GS}}(t)|\psi_\mathrm{GS}\rangle+|\psi_\mathrm{exc}(t)\rangle\), where \(\langle\psi_\mathrm{GS}|\psi_\mathrm{exc}(t)\rangle=0\) and \(|a_{\mathrm{GS}}(t)|^2=F(t)\). The corresponding interference contribution to the current is \(2\operatorname{Re}[a_{\mathrm{GS}}^*(t)\langle\psi_\mathrm{GS}|\hat{J}(t)|\psi_\mathrm{exc}(t)\rangle]\), which vanishes as the ground-state weight approaches zero. The observed correspondence therefore supports a temporal-gating mechanism involving depletion of this component, while allowing residual current within the excited-state sector. Because the evolution remains unitary, the fidelity reduction quantifies redistribution of wavefunction weight without requiring phenomenological dephasing. Together with the spectrograms, these results support the suppression of trailing-edge emission that exposes the rising-edge nonadiabatic blueshift.

\subsection{Universality with respect to the nearest-neighbor attraction}
\label{sec:1d hubbard}
\begin{figure*}[ht]
\includegraphics[width=.48\linewidth]{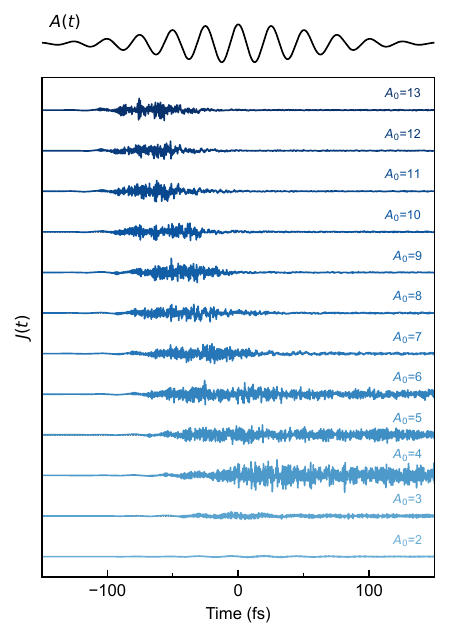}
\caption{\label{sfig:hhg 1d hubbard} Universality of gating in the pure one-dimensional Hubbard model. Time-dependent currents $J(t)$ in a half-filled 12-site periodic Hubbard chain ($U/t_h = 8$) driven by the vector potential $A(t)$ shown on top, for peak drive amplitudes $A_0 = 2$--$13$ (color-coded from light to dark). The traces are vertically offset for clarity.}
\end{figure*}

To confirm that this nonadiabatic gating is driven fundamentally by the Mott correlation $U$ rather than the nearest-neighbor interaction $V$, we simulated the charge dynamics in a pure Hubbard chain at half-filling by setting $V = 0$ while retaining $U = 8t_h$. The field-dependent evolution (Fig.~\ref{sfig:hhg 1d hubbard}) is essentially identical to that of the extended model (Fig.~\ref{fig:m4}): at weak field the current follows the pulse envelope and rings down only slowly afterwards, whereas above $A_0\simeq6$ the high-frequency oscillations are progressively quenched on the trailing edge, so that emission becomes increasingly restricted to the leading half of the pulse.

The persistence of this behavior in the absence of $V$ establishes that the nonadiabatic gating is not a consequence of the nearest-neighbor attraction but results from the on-site Coulomb correlation itself. The attraction $V$ renormalizes the optical gap and binds the photo-excited carriers, but it is the $U$-driven accumulation of these carriers and the attendant collapse of the ground-state fidelity that gates the emission. The nearest-neighbor attraction therefore only quantitatively tunes the threshold field and the carrier binding, leaving the gating qualitatively intact. Thus, many-body nonadiabatic gating is a generic property of strongly correlated Mott systems rather than a feature contingent on the specific range of the interaction strength.

\subsection{Generalization to two dimensions}
\label{sec:2d hubbard}
\begin{figure}[ht]
\includegraphics[width=.48\linewidth]{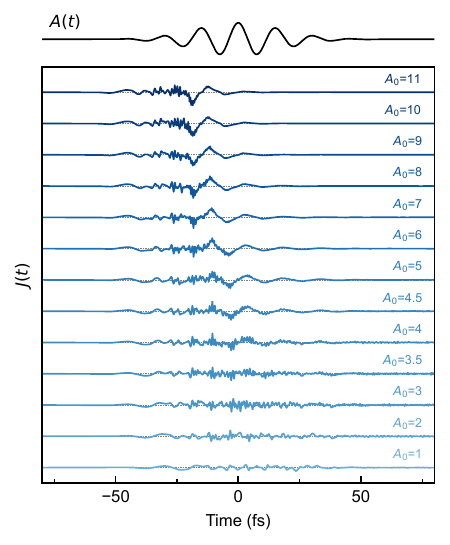}
\caption{\label{sfig:hhg 2d hubbard} Generalization of many-body gating to two dimensions. Time-dependent currents $J(t)$ in a half-filled $4\times4$-site periodic Hubbard cluster ($U/t_h = 8$) driven by the vector potential $A(t)$ along the nearest-neighbor direction shown on top, for peak drive amplitudes $A_0 = 1$--$11$ (color-coded from light to dark). The traces are vertically offset for clarity. Due to the high computational cost, the pulse is shorter than the duration in previous simulations on the Hubbard chain.}
\end{figure}

Finally, to test whether the mechanism is peculiar to one dimension---where spin--charge separation and the proximity to integrability of the Hubbard chain could in principle play a role---we repeated the simulation on a half-filled, $4\times4$-site periodic two-dimensional Hubbard cluster ($U=8t_h$, $V=0$, $t_h=0.35$~eV) driven along the nearest-neighbor direction ($\Omega=0.45t_h$), in accordance with experimental conditions. The resulting currents are shown in Fig.~\ref{sfig:hhg 2d hubbard}. Despite the markedly distinct single-particle spectrum and density of states \cite{arovas2022hubbard}, the field dependence mirrors the one-dimensional case: the current tracks the drive at low amplitude and is increasingly truncated on the trailing edge as $A_0$ grows.

This demonstrates that the field-driven coherence collapse responsible for the harmonic blueshift operates similarly in two-dimensional strongly correlated electron systems, and is therefore not a unique feature of one-dimensional Mott physics. The dimensional robustness is directly relevant to the experiment, where the same blueshift phenomenology is observed in the quasi-two-dimensional parent cuprate Sr$_2$CuO$_2$Cl$_2$ (Fig.~\ref{sfig:Sr2CuO2Cl2_HHG}) as in the quasi-one-dimensional chain compound Sr$_2$CuO$_3$. Taken together, Secs.~\ref{sec:1d hubbard} and~\ref{sec:2d hubbard} establish that many-body nonadiabatic gating is a universal consequence of a photo-induced electronic state reorganization in a driven Mott insulator, independent of the spatial dimensionality.

\clearpage

\section{Semiconductor Bloch equations simulations}
\label{sec:sbe}
To assess the limits of a weakly interacting Bloch-electron description, we investigated whether the observed highly blueshifted, non-integer harmonics can be explained using physically consistent parameters. We further examined whether artificially imposing a temporal gate as well as dynamic electronic structure within this framework can qualitatively reproduce the experimental wavelength-dependent crossover. 

\subsection{Two-band model}
\begin{figure*}[!hb]
\includegraphics[width=.5\linewidth]{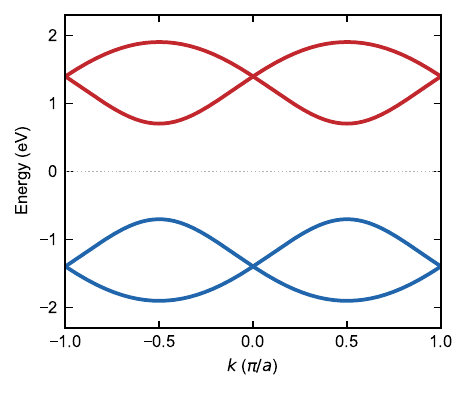}
\caption{\label{sfig:sbe band structure} Single-particle band basis for the SBEs simulations. Conduction (red) and valence (blue) bands, obtained from the exact Bethe-ansatz charge dispersion $\varepsilon(k)$ of the half-filled one-dimensional Hubbard model ($U/t_h = 8$).}
\end{figure*}

We solved the semiconductor Bloch equations (SBEs) on a basis derived from the exact charge excitation of the one-dimensional Hubbard model. The half-filled Hubbard model has a Bethe-ansatz holon–antiholon dispersion parametrized by the rapidity $k$ \cite{essler2005one}
\begin{align}
    p_h(k)&=p_d(k)+\pi=\frac{\pi}{2}-k-2\int_0^\infty\frac{d\omega}{\omega}\frac{\mathcal{J}_0(\omega)\sin{(\omega\sin{k})}}{1+\exp{\frac{U\omega}{2}}},
    \label{eq:holon exp1}\\
    \varepsilon_h(k)&=\varepsilon_d(k)=\frac{U}{2}+2\cos{k}+2\int_0^\infty\frac{d\omega}{\omega}\frac{\mathcal{J}_1(\omega)\cos{(\omega\sin{k})}e^{-\frac{U\omega}{4}}}{\cosh{\frac{U\omega}{4}}},
    \label{eq:holon exp2}
\end{align}
where $U$ is on-site Coulomb correlation in the unit of nearest-neighbor hopping $t_h$, and $\mathcal{J}_n$ is the Bessel function of the first kind. We neglect the nearest-neighbor attraction $V$ which has been shown not to affect the many-body nonadiabatic gating mechanism (Sec.~\ref{sec:1d hubbard}). Figure~\ref{sfig:sbe band structure} displays the single-particle band structure used for SBEs simulations with $U/t_h=8$ and $t_h=0.3$~eV, which give rise to a direct gap $E_g = 1.41$~eV that matches the absorption edge of Sr$_2$CuO$_3$ in experiments \cite{kim2008bound,ikeda2025high,baykusheva2026quantum}.

We propagate the two-band density matrix in the Houston frame to solve the nonlinear charge dynamics and the resulting HHs \cite{haug2009quantum,kira2011semiconductor}. For each momentum $q$, we retain the conduction-band population $n_q(t)$ and the interband coherence (microscopic polarization) $p_q(t)$, which obey the SBEs
\begin{align}
    \dot{p}_q &= -i\left[\varepsilon_g\big(q+A(t)\big)-i\gamma \right] p_q - i\Omega_q(t)\big(1 - 2 n_q\big),\\
    \dot{n}_q &= -2\mathrm{Im}\left[\Omega_q^{*}(t)p_q\right],
\end{align}
where $\varepsilon_g(k) = E_c(k) - E_v(k)$ is the holon gap, $\Omega_q(t) = E(t)d_{cv}\big(k(t)\big)$ is the Rabi coupling strength, and $d_{cv}$ is the interband transition dipole (taken constant, $\sim$one lattice constant). The Pauli-blocking factor $(1-2n_q)$ retains the full population feedback. The system starts from the unexcited ground state, and $\gamma=1/T_2$ is a phenomenological dephasing rate encapsulating all scattering beyond the coherent two-band dynamics; here we take the standard value $T_2 = T/4$ (with $T$ the carrier period) \cite{vampa2014theoretical,heide2022probingeh}, generalized in the following to a population-dependent rate.

The emitted current separates into interband and intraband contributions, which are tracked independently. The interband (recombination) current is the time derivative of the macroscopic polarization,
\begin{equation}
    j_{\mathrm{inter}}(t) = \frac{d}{dt}P(t), \qquad
P(t) = 2\mathrm{Re}[\int\frac{dq}{2\pi} d_{cv}^{*}\big(q+A(t)\big)p_q(t)],
\end{equation}
while the intraband current is the band-velocity-weighted population,
\begin{equation}
    j_{\mathrm{intra}}(t) = \int\frac{dq}{2\pi}\big[v_c\big(q+A(t)\big) - v_v\big(q+A(t)\big)\big]n_q(t),
\qquad v_{c,v}(k) = \frac{\partial E_{c,v}}{\partial k}.
\end{equation}
The total current is $j(t) = j_{\mathrm{inter}}(t) + j_{\mathrm{intra}}(t)$, and the HHG spectrum is $I(\omega)\propto\big|\int \dot j(t)e^{i\omega t}dt\big|^2$ (Hann-windowed), with the corresponding Gabor transform giving the time-frequency maps.

\begin{figure*}[ht]
\includegraphics{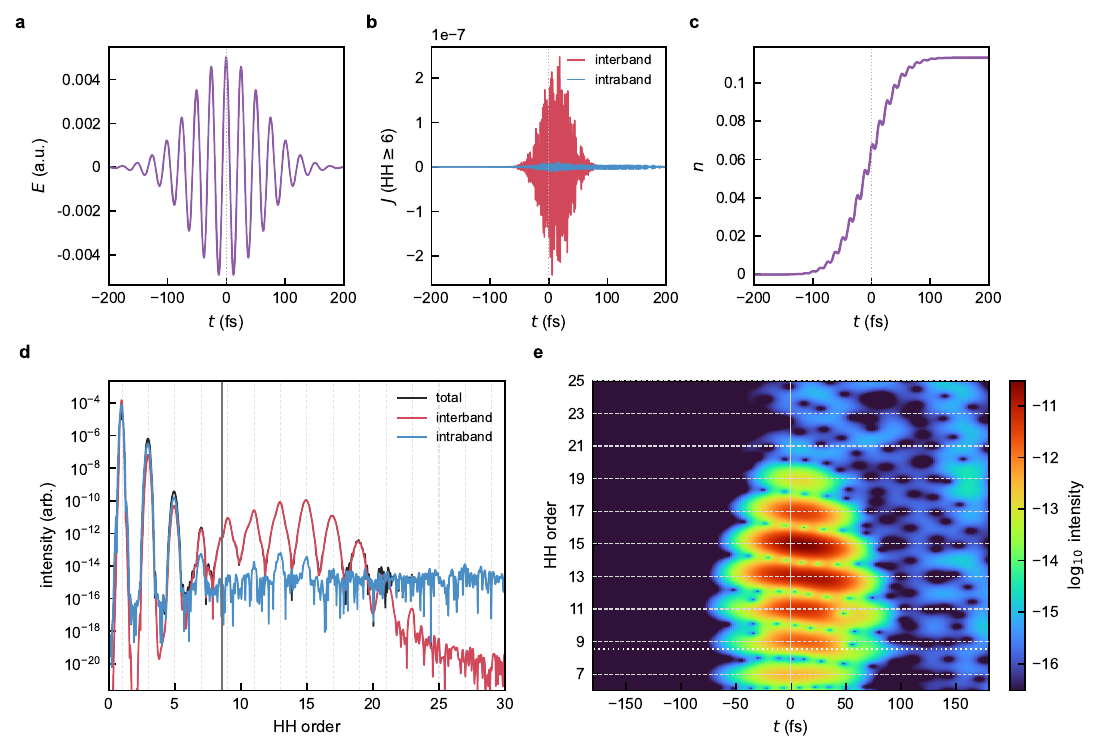}
\caption{\label{sfig:sbe constant t2} Numerical simulations of SBEs at constant dephasing $T_2 = T/4$. An MIR drive $\hbar\omega = 0.164$~eV, 100~fs intensity FWHM, with a peak field $E_0 = 5\times10^{-3}$ a.u., is used to generate HHs. (a) Driving electric field $E(t)$. (b) Interband (red) and intraband (blue) currents (high-pass filtered to retain harmonic orders $\geq 6$). (c) Carrier density dynamics. (d) HHG spectrum (black) decomposed into interband (red), intraband (blue) components. The vertical line marks the minimum charge gap and thin grey dotted lines denote the expected emission energies of odd-order HHs. (e) Gabor time-frequency map of the nonlinear current (HH7–25; dashed lines at odd orders, the dotted line at the minimum gap). The spectrogram is computed by Fourier transforming the current within a sliding Gaussian window of 25~fs FWHM.}
\end{figure*}

We apply a Gaussian-pulsed electric field with 100~fs intensity FWHM to mimic the experimental excitation condition throughout the simulation (Fig.~\ref{sfig:sbe constant t2}a). The equations are integrated with a fixed-step fourth-order Runge–Kutta scheme on a uniform time grid (step $dt=0.08$~fs, window $\pm 4$~FWHM), with the Brillouin zone sampled on $N_q = 512$ points over $[-\pi,\pi)$. To control the stiff, rapidly oscillating interband phase ($\varepsilon_g\sim$ eV), we factor it out analytically, writing $p_q(t) = y_q(t)\exp[-\Gamma_q(t)]$ with $\Gamma_q(t)=\int_{t_0}^{t}\big[\gamma + i\varepsilon_g(q+A(\tau))\big]d\tau$ and propagating the slowly varying $y_q$; this rotating-frame substitution preserves accuracy at large gap and short dephasing.

\subsection{HHG with constant dephasing}
We first take the standard solid-HHG dephasing $T_2=T/4$ and tune the field so that the photo-excited carrier density matches the experimental level (Fig.~\ref{sfig:sbe constant t2}c; $E_0 = 5\times10^{-3}$ a.u.). The single-particle response (Fig.~\ref{sfig:sbe constant t2}d) reproduces a clean odd-integer harmonic comb: the plateau (orders 7–19) is carried almost entirely by the interband current, while the intraband current contributes only the low orders. The time-domain HH map shows that the emission is centered on the pulse, with the rising-edge (blueshifted) and falling-edge (redshifted) contributions essentially almost canceling (Fig.~\ref{sfig:sbe constant t2}e). The result is therefore an unshifted, integer harmonic spectrum.

\subsection{HHG with excitation-induced dephasing}
\label{sec:sbe eid}
\begin{figure*}[!ht]
\includegraphics[width=\textwidth]{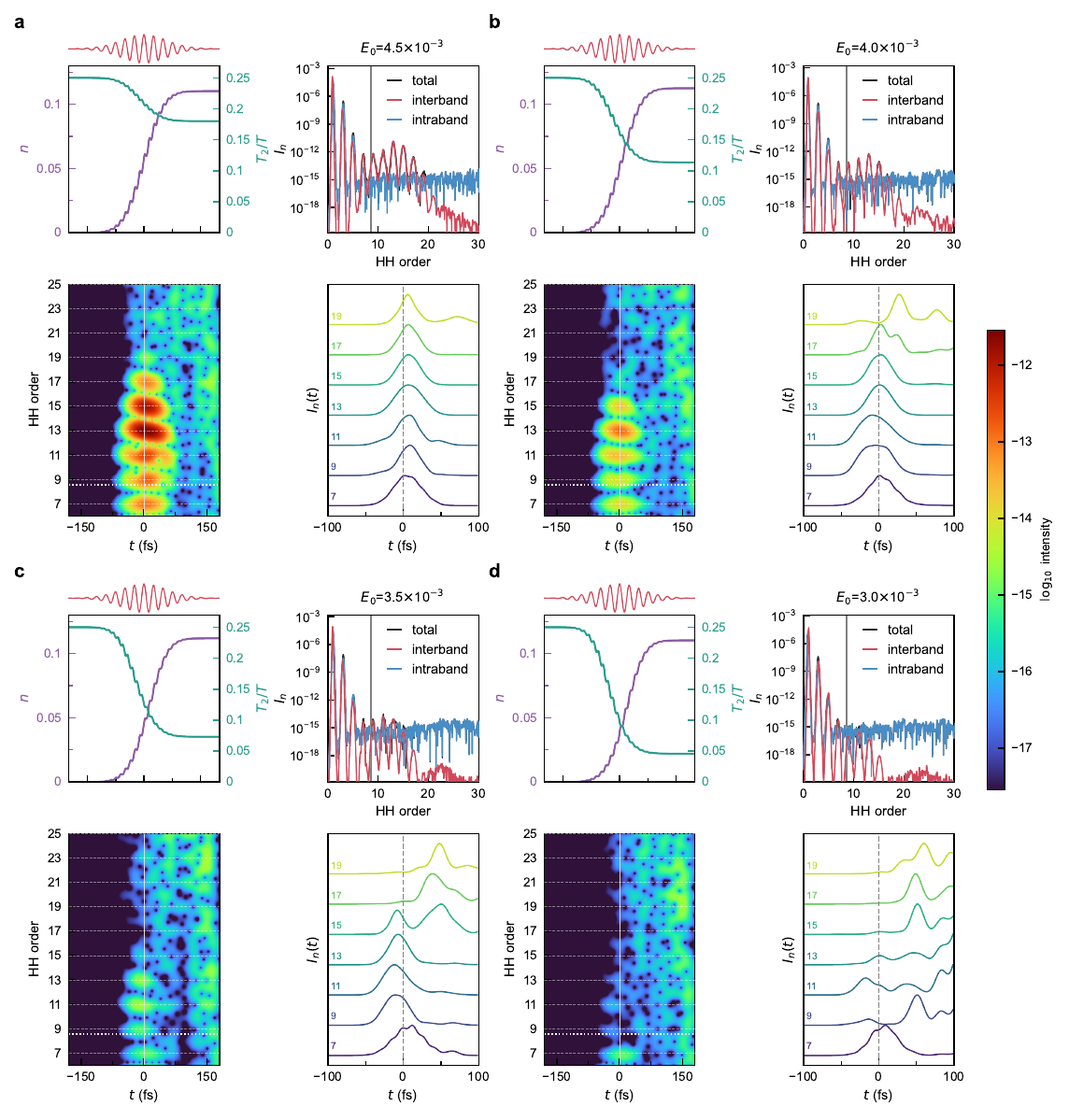}
\caption{\label{sfig:sbe alpha} Numerical simulations of SBEs with time-dependent dephasing $1/T_2 = 1/T_{2,0}+\alpha n$.  Within each sector: (top strip) the driving pulse $E(t)$ (red); (upper left) carrier density $n$ (purple, left axis) and the dynamic dephasing $T_2/T$ (teal, right axis) as a function of excitation time; (upper right) the HHG spectrum (black) with its interband (red) and intraband (blue) decomposition, with the minimum-gap order marked; (lower left) Gabor time-frequency map of the emission obtained by transformation within a sliding Gaussian window of 25~fs FWHM (HH7–25; dashed lines at odd orders, the dotted line at the minimum gap); (lower right) the per-odd-harmonic emission profile $I_n(t)$ (vertically offset for clarity; dashed line at $t=0$~fs). The peak electric field strengths are $E_0=4.5\times10^{-3}$ a.u. (a), $4\times10^{-3}$ a.u. (b), $3.5\times10^{-3}$ a.u. (c), $3\times10^{-3}$ a.u. (d).}
\end{figure*}

To break this temporal symmetry without ground-state depletion \cite{shin1999generation,shin2001nonadiabatic,chen2026two}, we set the coherence to decay faster as carriers accumulate, i.e., make the dephasing depend on the excited population. We therefore add a momentum-independent, excitation-induced dephasing (EID) \cite{nagai2023effect,guo2026dynamical},
\begin{equation}
    \frac{1}{T_2(t)} = \frac{1}{T_{2,0}} + \alpha N(t),
\end{equation}
where $N(t)$ is the total electron density and $\alpha$ the coupling (base $T_{2,0}=T/4$). We emphasize that this term is already beyond the single-particle description — a dephasing rate that responds to the carrier population is a carrier–carrier (many-body) effect on the mean-field level. Figure~\ref{sfig:sbe alpha} shows four runs at progressively weaker fields ($E_0 = 4.5,4,3.5,3\times10^{-3}$ a.u.); in each, $\alpha$ is tuned so the final carrier density again matches the experimental level (Fig.~\ref{sfig:carrier density}). Per sector we plot the pair density together with the dynamic $T_2/T$, the HHG spectrum and its interband/intraband decomposition, the time-frequency map of the nonlinear current, and the per-odd-harmonic emission profile $I_n(t)$ (obtained by integrating the time-frequency map between adjacent even orders). As the field is lowered at fixed carrier density, the required coupling rises steeply,
$$\hbar\alpha = 0.7,\ 2.2,\ 4.4,\ 8.5\ \text{eV} \quad (E_0 = 4.5 \to 3.0\times10^{-3}~\text{a.u.}),$$
strongly suppressing the coherence and progressively collapsing the harmonic emission. We note that the parameter set which reproduces a gated HH radiation towards the rising edge is $E_0=3.5\times10^{-3}$~a.u. and $\hbar\alpha=4.4$~eV, albeit with a significantly reduced plateau.

However, in any single-particle or mean-field treatment, the carrier dynamics is governed by the kinetic energy: a carrier that hops with amplitude $t_h$ cannot lose coherence much faster than this timescale. Equivalently, the SBEs are a weak-coupling expansion whose validity requires the residual carrier–carrier interaction to be a perturbation on the band motion ($V_{\rm eff}\ll W$, $V_{\rm eff}$ is effective carrier–carrier interaction matrix element and $W$ is the bandwidth). A genuinely weak-coupling EID is far smaller still, of order $t_h^2/W$, while the couplings demanded here reach $\hbar\alpha = 4.4$–$8.5$~eV and greatly exceed the bandwidth. Inverting Fermi's golden rule, $\hbar\alpha \sim |V|^2/W$, the implied scattering matrix element is $|V|\sim\sqrt{\hbar\alpha W}\sim U$, i.e. the bare on-site Coulomb interaction rather than a screened/kinetic one. A dephasing strength set by $U$ rather than by $t_h$ cannot be supplied by single-particle kinetics or mean-field scattering. Rather, it is by definition an interaction-driven, many-body effect, and at this magnitude it violates the very weak-coupling premise on which the SBEs rest.

Taken together (Figs.~\ref{sfig:sbe constant t2} and \ref{sfig:sbe alpha}), the single-particle SBEs on the correlated holon dispersion reproduce the integer harmonic comb but produce no blueshift at the experimental excitation with a constant dephasing rate. Reproducing the observed spectral evolution requires an EID whose coupling is of order — and beyond — the bandwidth and $U$, i.e. a bandwidth-scale, interaction-set scattering that lies outside the validity of any single-particle or mean-field description. Thus, we conclude that the nonadiabatic gating cannot be meaningfully approximated as independent particles undergoing phenomenological scattering. The truncation of the emission window, and the resulting blueshifted, non-integer harmonics, are intrinsically and irreducibly a many-body phenomenon driven by the collapse of the Mott-insulating ground state.

Related experimental observations have been interpreted in terms of EID. In two-color HHG measurements of Ca$_{0.4}$Sr$_{0.6}$CuO$_2$, higher-order signals peak at earlier relative delays and exhibit increasingly asymmetric profiles \cite{hatch2026sub}. These observations reveal an order-dependent two-color response consistent with dephasing-induced temporal reshaping. However, the reported delay-dependent yields do not constitute an independently calibrated reconstruction of the emission window relative to the driving-pulse envelope. Locating a gate-closure time would additionally require a calibrated temporal reference and characterization of the response to the finite-duration second-harmonic pulse. Thus, these measurements support dephasing-induced temporal selection but do not by themselves establish the specific many-body nonadiabatic gate considered here or its connection to harmonic blueshifts.

\subsection{HHG with electronic structure modification}
\label{sec:sbe gap reduction}
Photoexcitation can modify the electronic structure as well as the coherence lifetime in a correlated electron system. Following the phenomenological treatment of an evolving doublon--holon dispersion in~\cite{ikeda2025high}, we introduce a time-dependent effective Hubbard interaction into the two-band model. Here, instead of using an artificial ramping function, we describe the correlation reduction through carrier-dependent screening \cite{golez2019dynamics,baykusheva2022ultrafast},
\begin{equation}
    U(t)=U_0-\beta N(t),
\end{equation}
where $\beta=0.6$~eV per unit electron density. This prescription provides a minimal phenomenological description of the evolving electronic structure.

\begin{figure*}[htb]
\includegraphics{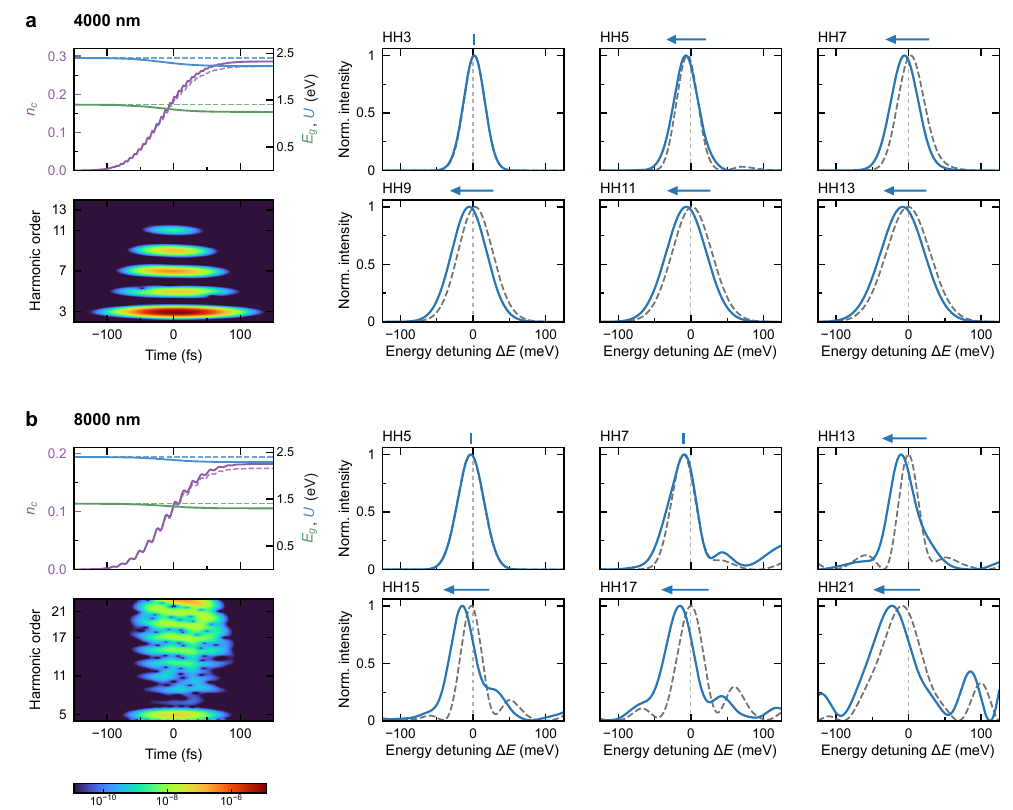}
\caption{\label{sfig:sbe tdU} Numerical simulations of SBEs with screening $U=U_0-\beta n$. Results for (a) 4000~nm and (b) 8000~nm excitation with strength $E_0=0.01$~a.u. In each wavelength group, the upper-left panel shows electron density $n_c$ (purple), Hubbard $U$ (blue), and the minimum bandgap $E_g$ (green). Solid and dashed curves represent the density-dependent-$U$ calculation and the fixed-$U$ reference, respectively. The lower-left panel shows the emission spectrogram for the former case, calculated from the time derivative of the total current using a Gaussian amplitude window with a 25~fs FWHM; both spectrograms share a logarithmic intensity scale. The remaining panels compare individual harmonic profiles for dynamic $U$ (blue solid) and fixed $U$ (gray dashed), each normalized to its own maximum within the displayed window. Arrows indicate the peak-shift direction as guides to the eye.}
\end{figure*}

During propagation, the instantaneous $U(t)$ determines the band dispersion through interpolation of precomputed dispersions, and the band velocities entering the current are updated consistently. The evolving bands therefore modify both the interband transition energies and the phase accumulated by the interband coherence between excitation and recombination. To examine the consequences of this feedback separately from EID, we keep $T_2=T/4$ constant and compare the resulting emission with a fixed-$U$ calculation under identical driving conditions.

Figure~\ref{sfig:sbe tdU} shows this comparison for 4000 and 8000~nm excitation at $E_0=0.01$~a.u., using a Gaussian pulse with a 100~fs intensity FWHM. The carrier buildup is accompanied by a decrease in $U$ and the minimum direct gap $E_g$. The normalized spectra show redward shifts of above-gap harmonics relative to their fixed-$U$ references at both wavelengths, whereas the subgap features remain nearly unchanged. Notably, the spectrograms show appreciable emission on both the rising and falling edges of the pulse, without any pronounced confinement to the rising edge. Thus, under these conditions, the evolving dispersion and its associated modification of the interband phase produce redshifts without requiring a strong temporal gate. This provides a reference for comparison with the preferential selection of rising-edge emission by EID, which favors a nonadiabatic blueshift in the regime considered below (Sec.~\ref{sec:sbe wavelength crossover}). The present comparison shows that redshifting through electronic-structure evolution and temporal gating are distinguishable effects that can coexist and compete. Thus, the observation of redshifted harmonics in Sr$_2$CuO$_3$ alone does not establish that the emission is temporally gated.


\subsection{Wavelength-tuned spectral shifts}
\label{sec:sbe wavelength crossover}
We now combine the carrier-dependent interaction introduced in Sec.~\ref{sec:sbe gap reduction} with the EID of Sec.~\ref{sec:sbe eid}. With the convention defined above, we use
\[
\frac{1}{T_2(t)}=\frac{1}{T_{2,0}}+\alpha N(t),\qquad
U(t)=U_0-\beta N(t),
\]
where we adopt the initial coherence time $T_{2,0}=T/4$ and \(\alpha T=200\). We retain the same $U(N)$ and 100~fs intensity FWHM at all wavelengths. Following the limitations discussed in Sec.~\ref{sec:sbe eid}, this deliberately strong dephasing is used as an imposed phenomenological gate to mimic the rapid suppression of coherent emission in the Hubbard-model calculations (Sec.~\ref{sec:ed}). The carrier population now controls both the coherence lifetime and the electronic structure, coupling their effects on the emitted spectrum.

Figure~\ref{sfig:sbe dynamics} illustrates the synchronized dynamics for a 10000~nm pulse with $E_0=0.01$~a.u. As carriers accumulate (Fig.~\ref{sfig:sbe dynamics}b), $T_2/T$ decreases from $1/4$ towards zero (Fig.~\ref{sfig:sbe dynamics}c). Concurrently, $U$ decreases from $2.4$ to approximately $2.11$~eV, and the minimum direct gap decreases from $1.41$ to roughly $1.14$~eV (Fig.~\ref{sfig:sbe dynamics}d). The high-pass-filtered current and spectrogram show that high-order emission is concentrated on the rising edge, with substantially weaker emission afterward (Fig.~\ref{sfig:sbe dynamics}e,f), resembling the temporal gate in Figs.~\ref{fig:m4} and~\ref{sfig:fidelity}. Meanwhile, the gap narrows during the active emission window, modifying the phase accumulated between excitation and recombination and introducing a competing trend toward redshift. The resulting spectral response therefore reflects the combined evolution of the emission window and the electronic structure. This coexistence motivates the wavelength-dependent comparison below, where changing the driving period alters the balance between temporal gating and gap renormalization.

The field-dependent spectra reveal how this competition evolves with driving wavelength (Fig.~\ref{sfig:sbe wavelength crossover}). At 10000~nm, the higher harmonic orders develop pronounced blueshifts with increasing field, consistent with the rising-edge temporal gate dominating their spectral evolution (Fig.~\ref{sfig:sbe wavelength crossover}a). At 4000~nm, the spectral weight evolves predominantly redward, consistent with gap renormalization outweighing the gating-induced blueshift (Fig.~\ref{sfig:sbe wavelength crossover}c). At the intermediate wavelength of 6500~nm, the harmonics exhibit nonmonotonic behavior: an initial redshift is followed by a recovery toward the integer harmonic energies and eventually a blueshift as the field increases (Fig.~\ref{sfig:sbe wavelength crossover}b). Because both effects depend on the evolving carrier population, these shifts reflect their coupled dynamics. Within this phenomenological framework, the calculations reproduce a crossover from predominantly redward to blueward spectral evolution with increasing wavelength, in qualitative agreement with the experimental trend (Fig.~\ref{fig:m3}).

\begin{supplementaryfigure}
\includegraphics{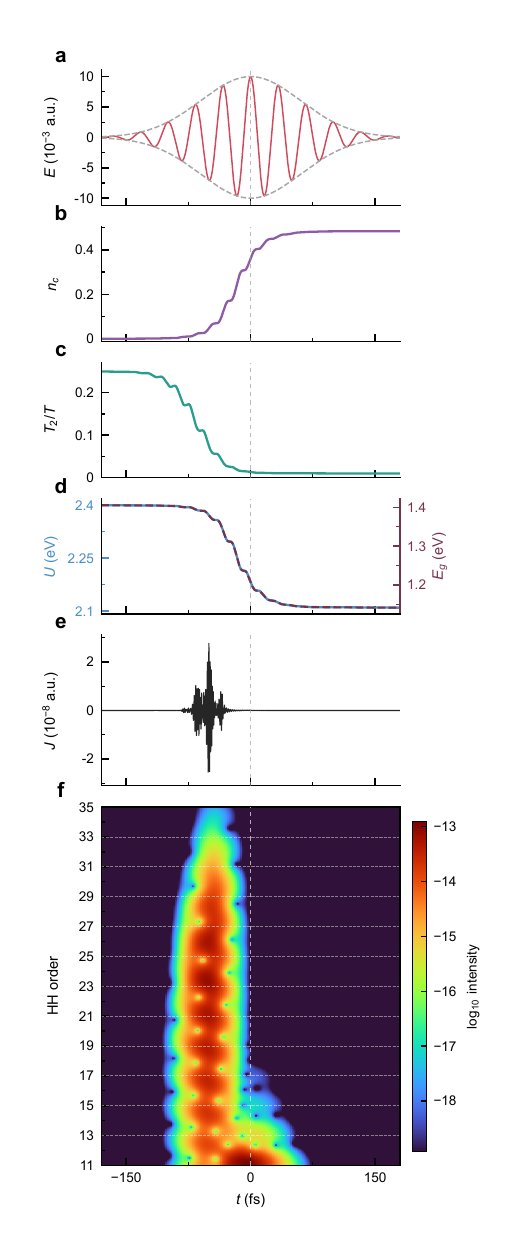}
\caption{\label{sfig:sbe dynamics} Time-domain dynamics in SBE simulations with EID $1/T_2 = 1/T_{2,0}+\alpha n$ and screening $U=U_0-\beta n$. (a) Driving electric field; grey dashed curves indicate its amplitude envelope. (b) Photoexcited electron density. (c) Normalized dephasing time $T_2/T$. (d) Hubbard interaction $U$ (blue, left axis) and instantaneous minimum direct gap $E_g$ (burgundy dashed curve, right axis). (e) High-harmonic current $J$, obtained by filtering the total current with a cosine high-pass transition from harmonic order 14. (f) Gabor spectrogram of the total-current derivative, calculated using a Gaussian amplitude window of 25~fs FWHM. Colors show logarithmic emission intensity, and horizontal dashed lines mark odd-integer harmonic orders. Vertical dashed lines indicate the pulse center. High-order emission is concentrated on the rising edge as the dephasing time decreases.}
\end{supplementaryfigure}

\begin{supplementaryfigure}
\includegraphics{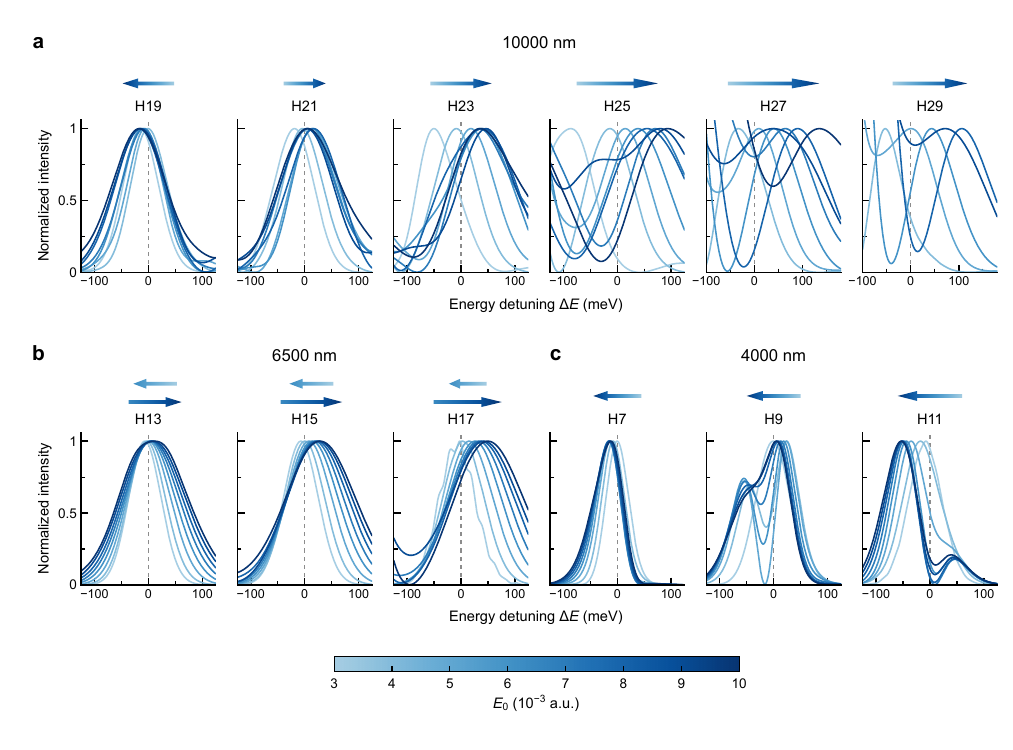}
\caption{\label{sfig:sbe wavelength crossover} Wavelength dependence of harmonic spectral profiles in SBE simulations with EID $1/T_2 = 1/T_{2,0}+\alpha n$ and screening $U=U_0-\beta n$. Calculated spectra near selected odd harmonic orders for driving wavelengths of (a) 10000, (b) 6500, and (c) 4000~nm. Darker blue denotes higher peak field, $E_0=0.003$–$0.010$~a.u. in steps of $0.001$~a.u. The horizontal coordinate is the detuning from the nominal integer harmonic energy; vertical dashed lines mark $\Delta E=0$. Each local profile is individually normalized and displayed on a linear intensity scale. Gradient arrows indicate the direction and approximate relative amplitude of spectral evolution.}
\end{supplementaryfigure}

\end{document}